\documentclass[twocolumn]{article}

\usepackage[
  letterpaper,
  top=1.8cm,
  bottom=1.8cm,
  left=1.6cm,
  right=1.6cm,
  columnsep=0.6cm
]{geometry}

\usepackage{sectsty}

\usepackage[T1]{fontenc}
\sectionfont{\large}
\subsectionfont{\small}
\usepackage[english]{babel}

\usepackage{csquotes}

\usepackage{xspace}

\usepackage{amsmath}
\usepackage{graphicx}
\usepackage[colorlinks=true, allcolors=blue]{hyperref}

\usepackage{balance}

\usepackage{amssymb}

\usepackage{fancyhdr}
\usepackage{graphicx}
\usepackage{dblfloatfix}   
\usepackage[font=small,labelfont=bf]{caption}

\usepackage{tcolorbox}

\title{\textbf{The power exhaust constrained SPARC separatrix operational space}}  

\author{
\textbf{    B. Lomanowski$^{1}$,
    T. Eich$^{2}$,
    J.D. Lore$^{1}$,
    J.-S. Park$^{1}$,
    T. Body$^{2}$,
    P. Stangeby$^{3,\dagger}$}
}
\date{\today}

\begin{document}









\newcommand{\Tet}{\ensuremath{T_{e\mathrm{,t}}}}
\newcommand{\alphat}{\ensuremath{\alpha_{\mathrm{t}}}}

\newcommand{\chiperp}{\ensuremath{\chi_\perp}}
\newcommand{\Pin}{\ensuremath{P_{\mathrm{in}}}}
\newcommand{\Psep}{\ensuremath{P_{\mathrm{sep}}}}
\newcommand{\Dperp}{\ensuremath{D_\perp}}
\newcommand{\qpar}{\ensuremath{q_\mathrm{\parallel,u}}}
\newcommand{\qdep}{\ensuremath{q_\mathrm{\perp}}}
\newcommand{\Gampar}{\ensuremath{\Gamma_{\mathrm{\parallel,t}}}}
\newcommand{\GamD}{\ensuremath{\Gamma_{\mathrm{D,core}}}}
\newcommand{\GamNe}{\ensuremath{\Gamma_{\mathrm{Ne}}}}
\newcommand{\qcyl}{\ensuremath{\hat{q}_{\mathrm{cyl}}}}
\newcommand{\Tesep}{\ensuremath{T_{e\mathrm{,sep}}}}
\newcommand{\nesep}{\ensuremath{n_{e\mathrm{,sep}}}}
\newcommand{\Teomp}{\ensuremath{T_{e\mathrm{,OMP}}}}
\newcommand{\TeompSH}{\ensuremath{T_{e\mathrm{,OMP}}^{\text{SH}}}}
\newcommand{\neomp}{\ensuremath{n_{e\mathrm{,OMP}}}}
\newcommand{\neompSH}{\ensuremath{n_{e\mathrm{,OMP}}^{\text{SH}}}}
\newcommand{\PLH}{\ensuremath{P_{\mathrm{L-H}}}}
\newcommand{\cNe}{\ensuremath{\langle c_{\mathrm{Ne}}\rangle_{\text{sep}}}}
\newcommand{\cz}{\ensuremath{c_{\mathrm{z}}}}
\newcommand{\Zeff}{\ensuremath{Z_{\mathrm{eff}}}}
\newcommand{\Dtwo}{\ensuremath{\mathrm{D_2}}}
\newcommand{\lamq}{\ensuremath{\lambda_{\mathrm{q}}}}
\newcommand{\lamn}{\ensuremath{\lambda_{\mathrm{n}}}}
\newcommand{\Idiv}{\ensuremath{I_{\mathrm{div}}}}
\newcommand{\fmom}{\ensuremath{(1-f_{\mathrm{mom}})}}
\newcommand{\fpwr}{\ensuremath{(1-f_{\mathrm{pwr}})}}
\newcommand{\nuedge}{\ensuremath{\nu_{e}^{*}}}
\newcommand{\kappaeFL}{\ensuremath{\langle \kappa_{e\parallel0\mathrm{,FL}}\rangle}}
\newcommand{\qparavg}{\ensuremath{\langle q_{e\parallel\text{cond}}\rangle}}

\twocolumn[

\maketitle

\begin{center}
\small
$^{1}$ Oak Ridge National Laboratory, Oak Ridge, TN 37831, USA \\
$^{2}$ Commonwealth Fusion Systems, Cambridge, MA 02139, USA \\
$^{3}$ University of Toronto Institute for Aerospace Studies, 4925 Dufferin St, Toronto M3H 5T6 Canada

\vspace{0.5em}
\footnotesize
$^{\dagger}$ Deceased before publication.
\end{center}

\begin{abstract}
In this work we extrapolate the separatrix operational space (SepOS) projections to SPARC and introduce detachment access criteria, thus formulating the combined power exhaust constrained SepOS (i.e., PE-SepOS) to evaluate integrated power exhaust solutions at scale. 
Through the interpretation of SPARC SOLPS-ITER datasets and foundational work already demonstrated in experiments we formulate an \textit{as simple as possible} description of the SOL net power and momentum losses in dissipative regimes to link the main power exhaust quantities (\qdep, \Tet, \cNe) with the SepOS parameters (\nesep, \Tesep, \Zeff, \alphat). Through this framework, we demonstrate the utility of a normalized PE-SepOS framework in identifying accessible operational points for given exhaustible \Psep{} requirements. In applying the PE-SepOS framework to project the SPARC operational space, we find inherent trade-offs, namely: i) accessing high impurity radiation scenarios leads to pronounced reductions in \nesep{} (e.g., 50\% reductions at \cNe=2\%) as a consequence of power limitation, and ii) given present understanding of access criteria to the quasi-continuous exhaust regime (QCE), a compromise between high radiative fraction and high density/neutral pressure is required for QCE access at sufficiently high density, high \alphat{} conditions, with the divertor dissipative regime transitioning to pronounced detachment. 
We further show that by establishing robust correlations between actuators \GamD, \GamNe, \Pin, and physics parameters \Tet{} and \cNe, the PE-SepOS framework can be used to inform experimental strategies in scaling up from low heating power to reactor-scale values of \Psep/R. Taking advantage of universal trends enabling projections of density and impurity seeding scans, the PE-SepOS thus provides a framework for mapping out the edge plasma operational space in a scalable manner, subject to validation during early SPARC operations with suitable divertor and edge plasma observables. 
\end{abstract}

\vspace{0.5cm}
\noindent\rule{\textwidth}{0.4pt}
\vspace{1em}
]

\section{Introduction}

The mitigation of excessive power and particle fluxes to plasma facing components (PFCs) remains a key challenge for magnetic confinement fusion. In reactor-scale tokamaks, the projected steady state unmitigated peak parallel heat flux densities in the scrape-off layer (SOL) upstream of the divertor targets, \qpar, are $\mathcal{O}(10~\mathrm{GW\,m^{-2}})$, whereas the limits on maximum deposited heat flux densities on PFCs imposed by material and engineering constraints are $\mathcal{O}(10~\mathrm{MW\,m^{-2}})$ based on empirical scalings of the SOL heat flux decay widths on existing tokamaks \cite{eich2011PRL,eich2013scaling,eich2020turbulence}. In high-Z metal wall devices compatible with reduced tritium retention, additional constraints arise from the need to minimize PFC erosion and high-Z impurity contamination of the core plasma due to physical sputtering from incident fuel  and impurity ion fluxes to the first wall. In the divertor this drives the need to operate at low enough steady state target electron temperature, \Tet $<$ 5-10 eV, to avoid excessive sputtering \cite{stangeby2018basic,van2013tungsten, matthews2011jet, brezinsek2015plasma}. Steady-state power exhaust solutions have successfully been implemented on today's devices by balancing high edge/divertor density typically achieved via strong gas fueling, and significant seeding of extrinsic impurities like $\mathrm{N_{2}}$, Ne and Ar to isotropically radiate excessive \qpar{} \cite{loarte2007chapter,loarte2011high,kallenbach2013impurity,reimold2015divertor,bernert2017power, henderson2021parameter}, also taking advantage of non-coronal transport effects that lead to more efficient radiative cooling than predicted by simple ionization balance estimates \cite{reinke2017heat,kallenbach2013impurity}. Extrapolating these demonstrated solutions to next-step devices with reactor-scale \qpar{} through high fidelity modeling \cite{kukushkin2011finalizing,pitts2019physics,lore2022high,lore2024evaluation} and empirical scalings of the required impurity concentration \cite{goldston2017new, reinke2017heat, kallenbach2019neutral, henderson2021parameter} leads to high projected radiative fraction ($f_{\text{rad}}$ $\sim$ 75-90\%) requirements and high divertor density and neutral pressures. The neutral compression and impurity \enquote*{enrichment} (i.e., how well the impurities are confined to the SOL-divertor region to avoid excessive core plasma dilution and radiative losses) remain active areas of research (e.g., \cite{kallenbach2024divertor}) that ultimately place limits on tolerable impurity content and edge densities from the core-edge integration perspective. Remaining uncertainties in extrapolating the near-SOL power widths to next-step devices results in a significant range of requirements, including the need to consider advanced divertor configurations (ADCs) \cite{kotschenreuther2013magnetic,cowley2023novel,moulton2024super} as risk mitigation for the most conservative power width estimates. ADC designs take advantage of divertor geometry and magnetic topology modifications to reduce the peak target heat fluxes relative to the conventional divertor designs. 

Compounding the steady-state power exhaust challenge is the typical presence of edge localized modes (ELMs) in high confinement (H-mode) operation. ELMs are characterized by periodic bursts of heat and particles into the SOL originating in the confined edge transport barrier (pedestal) region as a consequence of exceeding pedestal pressure gradient limits typically described by ideal MHD peeling-ballooning stability considerations \cite{snyder2011first}. Extrapolating the transient divertor heat and particle loads due to ELMs from existing devices based on empirical scaling of the ELM parallel energy fluence leads to an unfavorable scaling for reactor-scale devices that precludes operation in the so called large Type-I ELMy H-mode \cite{eich2017elm}, typically associated with high-performance pedestals. Focus has therefore shifted to the applicability and integration of small ELM or ELM-free scenarios \cite{dunne2025physics} with the aim of balancing core plasma performance demands (i.e., via sufficient pedestal pressures) while ensuring adequate PFC lifetimes by avoiding damaging large transients. 

Progress in mapping the tokamak operational space via the separatrix operational space framework (SepOS)  (\cite{eich2020turbulence,eich2021separatrix} and references therein) has enabled cross-machine comparisons of small/no-ELM regime access (e.g., QCE \cite{wolfrum2011characterization, faitsch2021broadening, harrer1292022, faitsch2023analysis, dunne2024quasi,dunne2025physics} or EDA H-mode \cite{greenwald1999characterization,miller2025determination}). The SepOS is derived from a data-driven approach on ASDEX Upgrade (AUG) through a classification of a large dataset of discharges using the separatrix density \nesep{} and temperature \Tesep, and further tested on Alcator C-Mod \cite{miller2025determination}. Building on earlier work by Rogers, Drake and Zeiler (RDZ) \cite{rogers1998phase} and Scott \cite{scott2005drift}, the SepOS framework comprises a concise set of equations describing the boundaries between L-modes and H-modes (LH), L-mode and H-mode density limits, as well as the ideal MHD boundary through considerations of turbulent growth rates and shear-flow suppression close to the separatrix. A key consideration in developing this heuristic framework has been trial and error in order to obtain agreement to the experimental data. Hence, this approach does not necessarily guarantee self-consistency, highlighting the need for careful validation beyond the existing database. Despite these limitations, a normalized version of the SepOS framework was recently developed by Eich \cite{eich2025separatrix} for AUG, Alcator C-Mod and SPARC\textsuperscript{\textregistered}\footnote{SPARC\textsuperscript{\textregistered} is a trademark of Commonwealth Fusion Systems\textsuperscript{\textregistered}} through a normalization of the LH SepOS boundary. This technique has enabled the projection of the AUG dataset to SPARC, and initial estimates of QCE access conditions. Projections currently rely on applying the empirical threshold value of the \alphat{} turbulence parameter, with \alphat{} $>$ 0.55 describing access to the QCE regime reasonably well in AUG. Global pedestal stability analysis in combination with more local stability calculations at the bottom of the pedestal near the separatrix have so far recovered the observed \alphat{} empirical QCE access scaling \cite{dunne2024quasi}.

Whereas steady state power exhaust solutions leveraging atomic/molecular physics loss channels and divertor topology/geometry optimization are in an advanced state of development on present day devices, the integration of an ELM-free or small-ELM scenario with detached divertor conditions  and acceptable plasma performance remains to be demonstrated at reactor-scale \qpar. The SPARC tokamak \cite{creely2020overview,creely2023sparc,rodriguez2024core,kuang2020divertor,lore2024evaluation, hughes2025high} currently in construction will be the first-of-its-kind device where such an integrated scenario will not only be attempted at scale, but will be critical to achieving its Q$>$1 mission while demonstrating PFC survivability. 

In this contribution we address the need to develop an integrated predictive capability combining SPARC SepOS projections for QCE/EDA H-mode \cite{eich2025separatrix} with power exhaust constraints. The main aim is to evaluate the range of accessible operational space for Q$>$1 scenarios that avoid damaging Type-I ELMs and accommodate sufficient steady-state mitigation of target heat fluxes and tungsten PFC erosion. We utilize high fidelity SOLPS-ITER \cite{wiesen2015new} datasets expanded from initial SPARC operational space studies \cite{lore2024evaluation,park2024actuator} with the SepOS H-mode boundaries and QCE access conditions extrapolated from AUG. In doing so, we quantify the strong link between the separatrix plasma parameters and the divertor detachment state moderated via gas fueling and Ne seeding. By further utilizing a simplified description of divertor dissipation which enables normalization for \Psep{} and SOL power width assumptions in the modeling, we formulate the power exhaust constrained SepOS (i.e., PE-SepOS) in order to evaluate the tradeoffs in the power exhaust solutions on SPARC at reactor-scale values of \qpar. We consider the balance in accessing the projected QCE regime while ensuring sufficient Ne impurity concentration for mitigating peak heat fluxes in the divertor to acceptable levels. Finally, we show that a key feature of the PE-SepOS framework is the self-similar trends relating the key upstream and target plasma parameters to the outer target electron temperature, \Tet, which we consider to be the main parameter describing dissipative divertor behavior.

\section{\label{sec:level1}Methodology}

\begin{figure}
    \centering
    \includegraphics[width=0.65\linewidth]{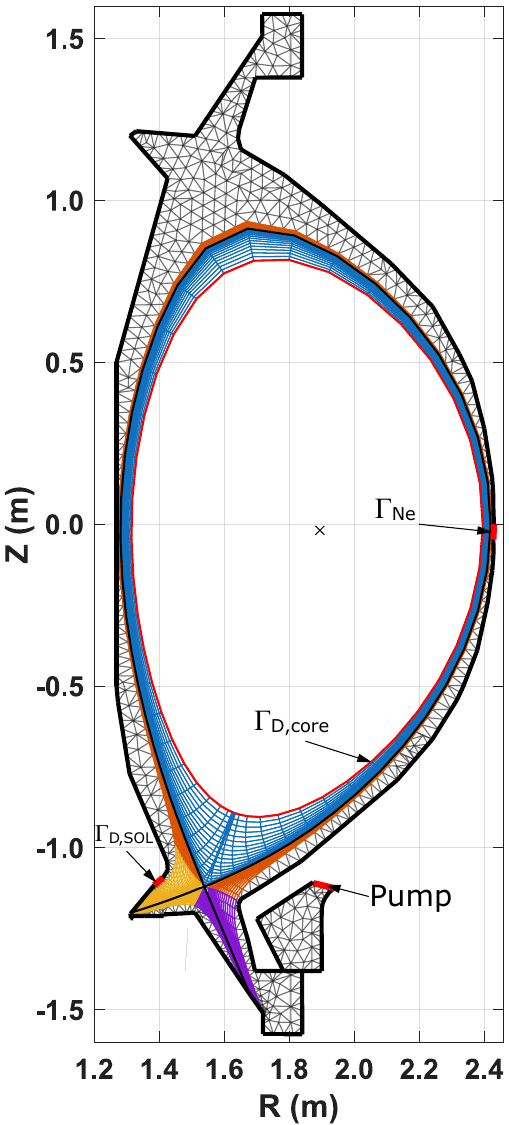}
    \caption{SOLPS computational mesh corresponding to a SPARC lower single null plasma shape with outer divertor strike point on the \enquote*{horizontal} divertor target, and SPARC V2y first wall geometry. }
    \label{fig:solps_grid}
\end{figure}

The SPARC SOLPS simulation dataset includes steady-state scans of deuterium fueling, heating power (\Pin{} $\approx$ \Psep{}$=$ 6 - 29 MW) and poloidally averaged Ne concentration at the separatrix, \cNe{} $<$ 2\% for a biased single-null ($dR_{\mathrm{sep}} =$ -2 cm) topology shown in figure \ref{fig:solps_grid}, with similar input settings as in earlier results presented in \cite{lore2024evaluation}. The density scans are actuated by a core particle source, \GamD, primarily due to numerical stability, with the range of edge densities exceeding in all cases considered here densities at which the previously reported HOCI (hot outer cold inner) / HICO (hot inner cold outer) bifurcations are present due to thermoelectric current-driven instabilities \cite{lore2024evaluation}. Fueling from the inner SOL is also considered in Section \ref{sec:at_vs_Tet}. Ne seeding is introduced at the outer midplane (OMP) in all cases. While the Ne seeding location is not expected to have a significant impact on the plasma solution, sensitivity to this input parameter has so far not been carried out. The analysis focus is on the peak flux tube in the outer SOL, which is typically the operationally limiting flux tube exhibiting peak heat flux densities at the outer target in attached divertor conditions. In the SOLPS computational grid, this corresponds to the third flux tube in the SOL, 0.1 mm from the separatrix at the OMP, 1.9 mm away from the outer divertor strike point.

Two values of the anomalous radial heat transport coefficients in the SOL were selected for the study: \chiperp{} = 0.05 $\mathrm{m^2}$/s, representing H-mode-like \lamq{} $\approx$ 0.3-0.8 mm depending on density (see Fig. 3 in \cite{lore2024evaluation}) for the 12.2 T, 8.7 MA scenario, and \chiperp = 0.10 $\mathrm{m^2}$/s representing L-mode-like \lamq{} values for the 12.2 T, 8.7 MA scenario, consistent with about a factor of two to three difference in L and H-mode near-SOL decay widths typically observed in experiments \cite{sieglin2017density}. These values are consistent with the updated near-SOL power width scalings from Eich \cite{eich2020turbulence}, evaluated in SOLPS based on exponential fits either directly from \qpar{} near-SOL profiles at the X-point and mapped to the outer mid-plane, or from profiles of the electron temperature and $\lambda_q=2/7 \lambda_T$ obtained from the Spitzer-H{\"a}rm parallel electron conduction power balance model \cite{stangeby2000plasma,eich2020turbulence}.  \Dperp/\chiperp $=$ 10 is held constant in all simulations for consistency with the initial SPARC SOLPS datasets from \cite{lore2024evaluation}. Additional details on the parametric scans of \Dperp{} and \chiperp{} and comparison to the updated Eich near-SOL decay width scalings are provided in \cite{park2026DsChis}.

For the higher heating power scans that are above the projected LH SepOS boundary (see figure \ref{fig:sepos_12Tvs8T_rollover}), the difference in the two \chiperp{} values is also consistent with the difference between the full field 12.2 T, 8.7 MA scenario, and the 8 T, 5.7 MA reduced field scenario with more favorable H-mode access (see \cite{hughes2025high} for details), and a wider projected \lamq$\propto B_{\text{pol}}^{-1}$. Hence, although our focus is on the 12.2 T scenario, the datasets are also informative for the 8 T scenario scoping, as will be shown in Section \ref{SepOS}.

\section{Simplified description of divertor dissipation}
\begin{figure}
    \centering
    \includegraphics[width=0.95\linewidth]{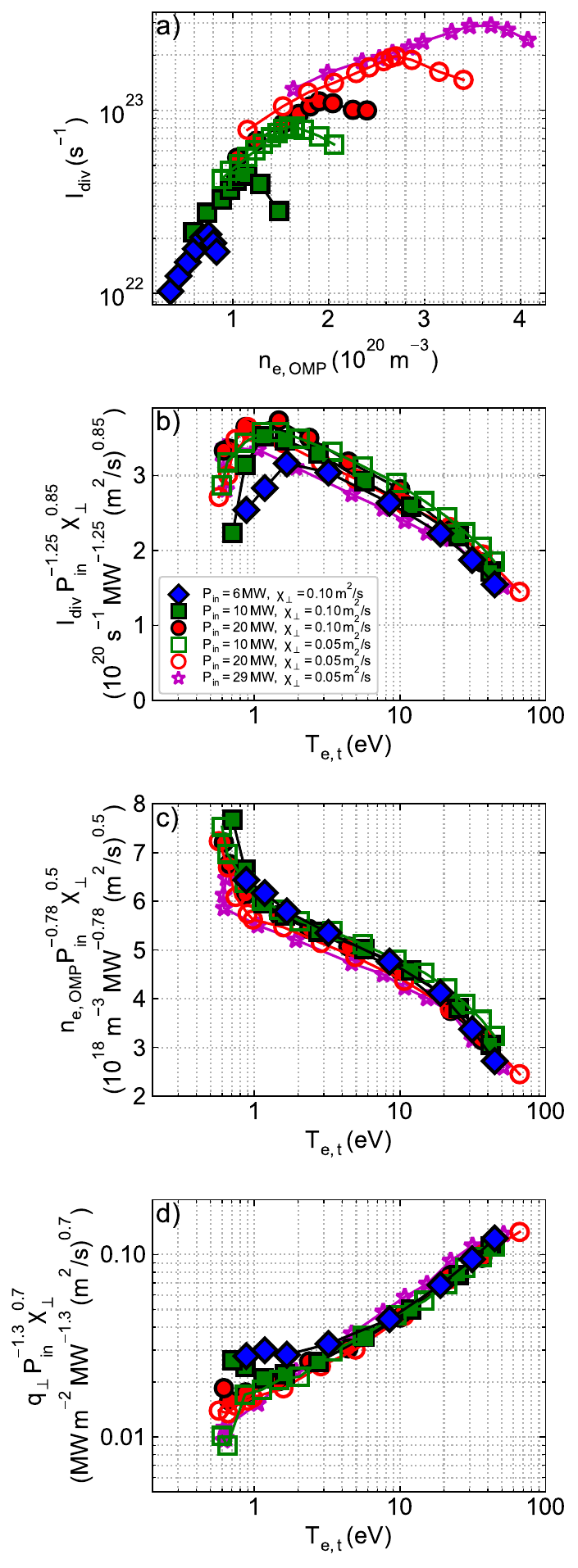}
    \caption{Dependence of a) normalized \Idiv on \neomp; b-d) normalized \Idiv, \neomp and \qdep on \Tet for unseeded density scans at different values of \Pin and \chiperp. \Idiv is integrated over the outer SOL divertor target radial profile, whereas all other quantities are on the peak flux tube.}
    \label{fig:Idiv_neomp_Tet}
\end{figure}

The range of \Psep{} and \GamD{} captured in the SOLPS dataset yields a considerable range of SOL density and outer divertor target conditions. Figure \ref{fig:Idiv_neomp_Tet}.a shows the range of integrated particle flux at the outer target, $I_{\mathrm{div}} = \int_{\mathrm{target}} \Gamma_{\parallel} \sin(\alpha) \, dA$, mapped to the upstream density, \neomp $\approx$ \nesep \footnote{We evaluate the OMP upstream quantities \neomp{} and \Teomp{} at the peak flux tube in the SOL throughout the paper to directly connect to the outer divertor target. \Tesep$\approx$\Teomp{} and \nesep$\approx$\neomp{} given the close proximity of the peak flux tube to the separatrix.}, exhibiting the characteristic \Idiv{} rise, up to the particle detachment rollover, followed by an \Idiv{} reduction at higher densities and deeper detachment. For simplicity and convenience, we use the \Idiv{} rollover point as a reference dividing the unseeded partially detached regime at lower \neomp{} values, and the unseeded pronounced detached regime at higher \neomp{} beyond the rollover point. This is consistent with the typical particle detachment interpretation in tokamak experiments. 

Given the similarities in the \Idiv{} rollover trends, it is convenient to map the density scans (i.e., \GamD{} scans) to the target electron temperature, \Tet, since we consider this to be the main physics parameter that characterizes divertor detachment. The utility of mapping detachment trends to \Tet{} has been described in multi-machine reduced models for plasma detachment \cite{stangeby2018basic,Stangeby_2020_Part1,Stangeby_2020_Part2}, as well as from JET experiments and interpretive modeling \cite{lomanowski2019spectroscopic, lomanowski2020interpretation, lomanowski2022experimental, lomanowski2023variation,lomanowski2023parameter}. These studies provide a numerical and experimental basis for the robust correlations of upstream quantities like \neomp{} with \Tet, as well as the strong dependence of the volumetric momentum, $(1-f_{\mathrm{mom}})=p_{\mathrm{t}}^{\mathrm{tot}}/p_{\mathrm{u}}^{\mathrm{tot}}$, and power loss, $(1-f_{\mathrm{\mathrm{pwr}}})=(q_{\mathrm{\parallel,t}}R_{\mathrm{t}})/(q_{\mathrm{\parallel,u}}R_{\mathrm{u}}) \approx (7.5kT_{e\mathrm{,t}}\Gamma_{\mathrm{\parallel,t}}R_{\mathrm{t}})/(q_{\mathrm{\parallel,u}}R_{\mathrm{u}})$, factors on \Tet{} (where the \enquote*{u} and \enquote*{t} subscripts refer to the upstream and divertor target quantities on the peak flux tube, respectively). Although \Tet{} is not controlled directly in experiment, and doesn't directly describe the dissipative loss mechanisms, its utility as the main detachment physics parameter lies in unifying disparate trends and reducing the dimensionality of the net dissipative divertor properties to an \textit{as simple as possible} description.

Recasting the unseeded datasets shown in \ref{fig:Idiv_neomp_Tet}.a in terms of \Tet{} at the peak flux tube, and further deriving normalization factors for \Pin $\approx$\Psep{} and \chiperp,  figure \ref{fig:Idiv_neomp_Tet}.b shows the self-similar \Idiv\Pin$^{-1.25}$\chiperp$^{0.85}$ trends. We extend similar normalizations to other quantities of interest, including  \neomp\Pin$^{-0.78}$\chiperp$^{0.5}$ and \qdep\Pin$^{-1.3}$\chiperp$^{0.7}$ in figure \ref{fig:Idiv_neomp_Tet}.c and d, where \qdep{} is the deposited heat flux comprising conducted and convected plasma flux, heat flux from thermoelectric current, potential energy through recombination, as well as the net flux of neutral atoms and molecules and radiation, as in \cite{lore2024evaluation}. The derived normalization exponents for \Pin{} and \chiperp{} are obtained manually and are informed, approximately, from the two point model (2PM) equations obtained from rearranging equations 15b, 17b and 42 in \cite{stangeby2018basic},  
\begin{equation}
    n_{e\mathrm{,OMP}} \propto \frac{q_{\mathrm{\parallel,u}}^{5/7}}{T_{e\mathrm{,t}}^{1/2} L_{\mathrm{\parallel}}^{2/7}} \left[ \frac{(1-f_{\mathrm{pwr}})}{(1-f_{\mathrm{mom}})} \right],
    \label{eq:neomp}    
\end{equation}

\begin{equation}
    \Gamma_{\mathrm{\parallel,t}} \propto \frac{q_{\mathrm{\parallel,u}}}{T_{e\mathrm{,t}}}
    (1-f_{\mathrm{pwr}}),   
    \label{eq:Gampart}    
\end{equation} and by normalizing with respect to $q_{\mathrm{\parallel,u}} \propto P_{\mathrm{sep}} / \lambda_{\mathrm{q}}(\chi_{\mathrm{\perp}})$. We note that a simple analytic expression for \lamq(\chiperp) is not available and since \Pin{} and \chiperp{} are inputs directly controlled by the user rather than \qpar{}, this approach allows for more convenient formatting of the code results. 

\begin{figure}
    \centering
    \includegraphics[width=0.9\linewidth]{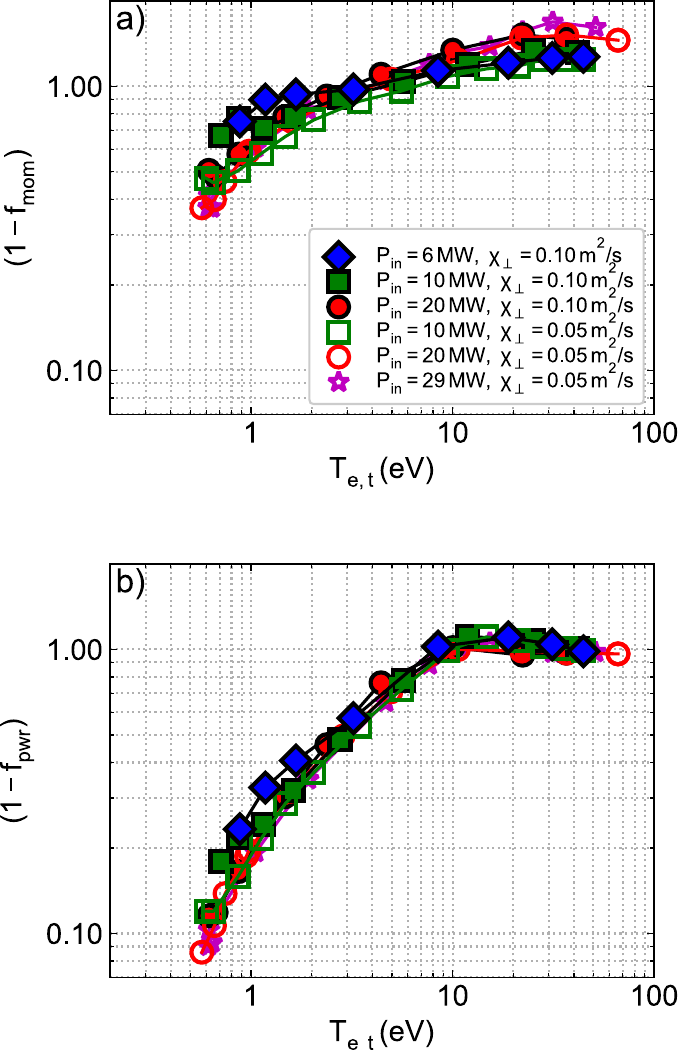}
    \caption{Dependence of a) total momentum and b) power volumetric loss factors on \Tet{} at the peak flux tube for unseeded density scans at different values of \Pin{} and \chiperp. The upstream location is at the OMP for momentum losses, and at the X-point for power losses to avoid the influence of \Psep{} entering the SOL upstream of the X-point.}
    \label{fig:fmomfpwr}
\end{figure}

The strong \fmom{} and \fpwr{} volumetric loss factor dependence on \Tet{} is shown in figure \ref{fig:fmomfpwr}, where the upstream total pressure, $p_{\mathrm{u}}^{\mathrm{tot}}$, location is taken at the OMP, and the $q_{\mathrm{\parallel,u}}$ location is taken at the X-point to avoid the influence of \Psep{} entering the SOL upstream of the X-point. The physical mechanism for the strong \Tet{} dependence on \fmom{}  has been attributed to momentum losses through plasma-neutral interactions close to the target. Detailed descriptions of the momentum loss channels are described in \cite{stangeby2018basic,kotov2009two} and references therein. For example, in SOLPS simulations for ITER these losses were shown to occur in the plasma volume 1 cm away from the target \cite{park2024impact}. Additionally, in the numerical dataset synthesis in \cite{stangeby2018basic} the volumetric momentum and cooling losses were found to be strongly correlated, as long as  $p_{\mathrm{u}}^{\mathrm{tot}}/q_{\mathrm{\parallel,u}}$ remains approximately constant, which is indeed the case in the unseeded SPARC simulations. Note that the thermoelectric current driven asymmetries and associated hysteresis described in \cite{lore2024evaluation} (i.e., the transition from HOCI to HICO regimes) are not present for the range of simulations shown in figures \ref{fig:Idiv_neomp_Tet} and \ref{fig:fmomfpwr} since even the lowest density cases in the SOLPS datasets considered for this study lie above the  hysteresis phase space.   

In addition to the numerical studies discussed above, an experimental basis for the strong correlations of \neomp{} and \Gampar{} with \Tet{} has also been established on JET with the W-Be wall \cite{lomanowski2022experimental,lomanowski2023parameter, lomanowski2023variation}. Utilizing passive divertor spectroscopy techniques to estimate \Tet{} \cite{lomanowski2020interpretation}, the \Idiv{} and \neomp{} dependence on divertor geometry-specific fueling and pumping characteristics was removed when correlated to \Tet, unifying disparate configuration specific trends \cite{lomanowski2022experimental}. Similarly, detailed measurements of \fmom{} and \fpwr{} enabled by divertor spectroscopy showed good agreement with interpretive EDGE2D-EIRENE modelling for unseeded conditions \cite{lomanowski2023parameter}, further validating the utility of \Tet{} in characterizing detachment.

With the above \Tet{} mappings and 2PM-informed normalizations for \Pin{} and \chiperp, the density scan trends in figures \ref{fig:Idiv_neomp_Tet} and \ref{fig:fmomfpwr} exhibit a simple, self-similar form that provides the basis for linking the key divertor quantities to the SepOS framework.

\section{\label{SepOS}Separatrix operational space projections}

\begin{figure*}[t]
    \centering
    \includegraphics[width=0.8\linewidth]{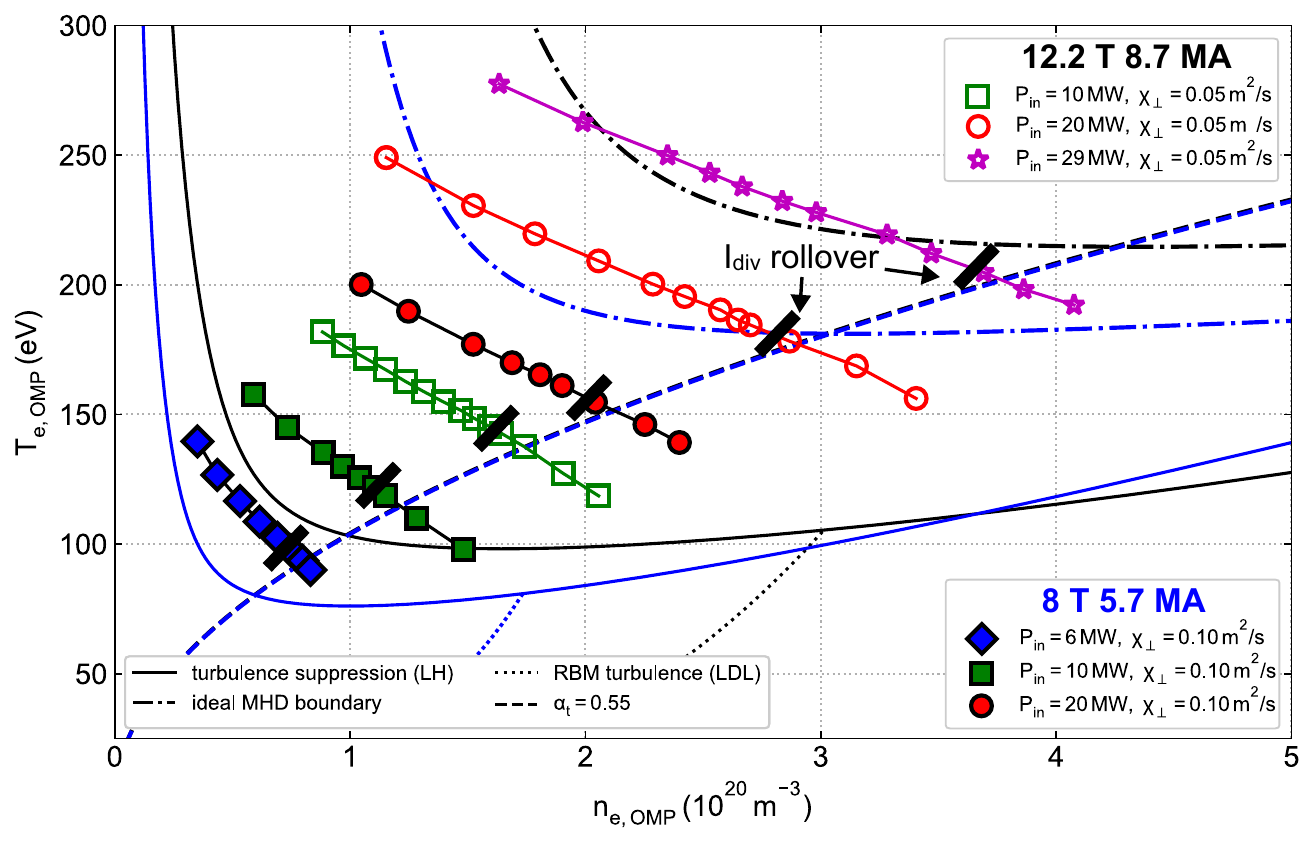}
    \caption{The projected SPARC separatrix operational space boundaries and \alphat$=0.55$ contours for the full field $B_{\text{t}}=$12.2 T, $I_{\text{p}}=$8.7 MA and reduced field $B_{\text{t}}=$8 T, $I_{\text{p}}=$5.7 MA scenarios. Markers correspond to unseeded density scans at different \Pin{} and \chiperp{} values. The \Idiv{} rollover point for each scan is indicated.}
    \label{fig:sepos_12Tvs8T_rollover}
\end{figure*}

Using the SepOS operational boundaries for the LH transition (i.e., turbulence suppression condition), ideal-MHD limit and the L-mode density limit (i.e., resistive ballooning mode (RBM) turbulence condition) extrapolated from AUG (\cite{eich2021separatrix,eich2025separatrix} and references therein, made available in the \href{https://github.com/cfs-energy/cfspopcon}{cfsPOPCON} package \cite{tom_body_2024_13820268}) the projected SPARC SepOS with unseeded SOLPS density scans is shown in figure \ref{fig:sepos_12Tvs8T_rollover}. \Teomp{} and \neomp{} are the upstream OMP values of electron temperature and density at the peak flux-tube, chosen instead of the values at the separatrix to maintain consistency with the 2PM-like upstream-downstream comparisons along the same flux tube. The differences between \Tesep, \nesep{} and \Teomp{} and \neomp{} on the peak flux-tube are negligible since the peak flux tube is less than 0.1 mm away from the separatrix at the OMP, and hence significantly below one $\lambda_{\mathrm{q}}$. The density scans span the L-mode and H-mode regions of the SepOS, with a strong \Teomp{} and \neomp{} dependence on \Pin{} $\approx$ \Psep{} and \chiperp.  As noted in \cite{eich2025separatrix}, the SepOS LH boundary extrapolations from AUG to SPARC should not be over-interpreted since the impact of the ion heat flux channel and ion to electron temperature ratios have not been addressed thus far. More detailed predictions of SPARC H-mode access are given in \cite{hughes2025high} using both the up-to-date International Tokamak Physics Activity (ITPA) empirical scalings \cite{delabie2026empirical}, as well as the ion heat flux threshold scalings from \cite{ryter2014experimental}. 

The equivalent SPARC SepOS projection for the reduced field 8 T, 5.7 MA scenario is also shown in figure \ref{fig:sepos_12Tvs8T_rollover} relative to the full field 12.2 T, 8.7 MA scenario. We can expect \lamq{} to be approximately 8.7/5.7$\approx$1.53 wider for the 8 T, 5.7 MA scenario based on \lamq$\propto B_{\text{{pol}}}^{-1}$ scaling. We can therefore consider the \chiperp=$0.1$ $\mathrm{m^2/s}$ density scans to inform the reduced field SepOS projections relative to the \chiperp=$0.05$ $\mathrm{m^2/s}$ scans which are more representative of the full field scenarios. In both cases, the \Pin$=$10 MW scans are above the SepOS LH boundary, more consistent with the projections for SPARC in \cite{hughes2025high} based on the critical ion heat flux scalings in \cite{schmidtmayr2018investigation} rather than from the recently updated ITPA LH threshold power scalings \cite{delabie2026empirical}. Regardless of the remaining uncertainty in the LH threshold power, we proceed simply by treating the SepOS LH boundary as illustrative and subject to further refinement.

\subsection{\label{sec:at_vs_Tet}Correlation between the target temperature and $\alpha_{\text{t}}$}

\begin{figure}
    \centering
    \includegraphics[width=.85\linewidth]{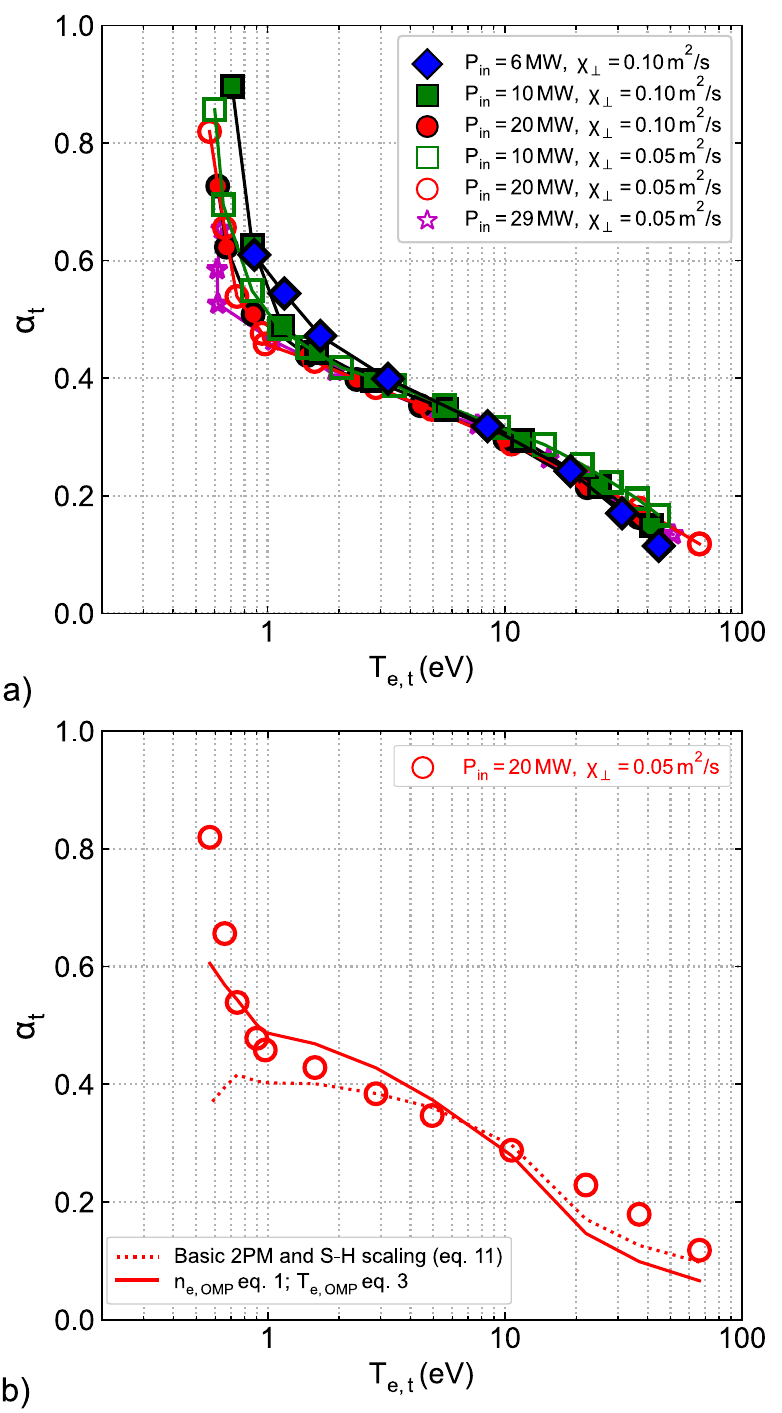}
    \caption{a) Dependence of \alphat{} on \Tet{} for the unseeded density scans at different values of \Pin{} and \chiperp. b) Different model assumptions for estimating \alphat{} relative to SOLPS estimates using \neomp{} and \Teomp{} directly for the \Pin$=20$ MW, \chiperp=$0.05$ $\mathrm{m^2/s}$ density scan.}
    \label{fig:Tet_alphat}
\end{figure}

\begin{figure}
    \centering
    \includegraphics[width=.92\linewidth]{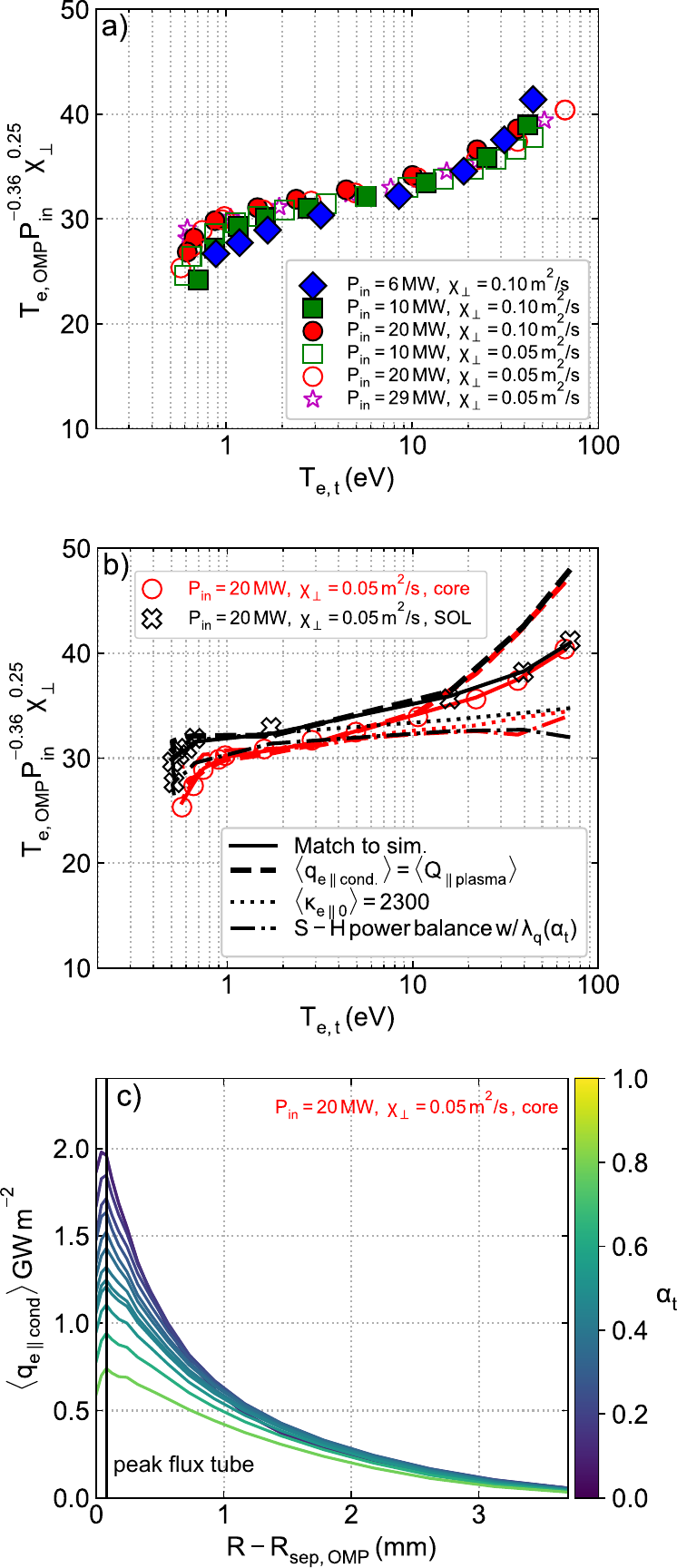}
    \caption{Dependence of a) normalized \Teomp{} on \Tet{} for the unseeded density scans at different \Pin{} and \chiperp{} values; b) comparison of the impact of fueling location (core vs inner SOL) on the normalized \Teomp{} for the density scans at \Pin$=20$ MW, \chiperp=$0.05$ $\mathrm{m^2/s}$ including different model assumptions for estimating the SOLPS output; c) \qparavg{} radial profiles for the density scan at \Pin$=20$ MW, \chiperp=$0.05$ $\mathrm{m^2/s}$, with the line color indicating the value of \alphat.}
    \label{fig:Teomp_breakdown}
\end{figure}

It is convenient to introduce the \Idiv{} rollover as a detachment reference point in figure \ref{fig:sepos_12Tvs8T_rollover}, as well as to add the \alphat{}$
     \propto\hat{q}_{\mathrm{cyl}}^{2} (n_{e\mathrm{,OMP}}/T_{e\mathrm{,OMP}}^{2}) Z_{\mathrm{eff}}
    \label{eq:placeholder_label}
=$ 0.55 empirical QCE threshold in order to interpret the SepOS \Pin{} and \chiperp{} dependence and link the degree of detachment to QCE access predictions. In estimating \alphat, we assumed a highly shaped scenario with plasma elongation $\kappa_{\mathrm{geo}}$=1.7 and triangularity $\delta$=0.5 using equations 4 and 5 in \cite{eich2020turbulence} for $\hat{q}_{\mathrm{cyl}}$. For these shaping parameters and the fixed \Dperp/\chiperp{} ratio, the \Idiv{} rollover point approximately intersects the \alphat=0.55 contour for all density scans, independent of \Pin{} and \chiperp. This independence emerges as a consequence of a strong correlation between \Tet{} and \alphat{} as more clearly shown in figure \ref{fig:Tet_alphat}.a. More specifically, the correlation is established between \Tet{} and $n_{e\mathrm{,OMP}}/T_{e\mathrm{,OMP}}^2$, since \qcyl{} and \Zeff{} are constant for the unseeded density scans. The strong dependence of \neomp{} on \Tet{} is already clear from figures \ref{fig:Idiv_neomp_Tet}.c, \ref{fig:fmomfpwr} and equation \ref{eq:neomp}, where \fmom{} and \fpwr{} can be described by $(1-f)\approx A(1-e^{-T_{e,\mathrm{t}}/b})^c$, where $A$, $b$, and $c$ are fit parameters \cite{stangeby2018basic}. However, it is not obvious why a significant decrease in \Teomp{} is observed with increasing \neomp{} in figure \ref{fig:sepos_12Tvs8T_rollover}. This is further explored in figure \ref{fig:Teomp_breakdown}.a which shows self-similar \Teomp\Pin$^{-0.36}$\chiperp$^{0.25}$ trends as a function of \Tet{} for all density scans, where the \Pin$^{-0.36}$ normalization is close to the expected \Teomp$\propto q_{e\parallel \text{cond}}^{2/7}$ Spitzer-H{\"a}rm scaling \cite{stangeby2000plasma}. Similar to the normalized trends mapped to \Tet{} in figures \ref{fig:Idiv_neomp_Tet} and \ref{fig:fmomfpwr}, there is little variation in the \Teomp{} trends over the range of \Pin{} and \chiperp, reinforcing the robust correlation of the key SOL-divertor parameters with \Tet. 

The significant decrease in \Teomp{} with increasing \neomp{} requires further analysis in order to understand the \alphat{} correlation with \Tet{} shown in figure \ref{fig:Tet_alphat}.a, since \alphat{} $\propto$ \Teomp$^{-2}$. We adopt the approach from \cite{moulton2024super} in using post-processed 2PM-like expressions for \Teomp{} to disentangle the main drivers. Starting with the first expression from Table 1 in \cite{moulton2024super},
\begin{equation}
    \label{eq:Tesep_sim_match}
    T_{e\mathrm{,OMP}}\approx\left(
        \frac{7}{2}
        \frac{
            \left\langle q_{e\parallel \mathrm{cond}} \right\rangle {L}_{\parallel}
        }{
            \left\langle \kappa_{e\parallel \mathrm{0,FL}} \right\rangle 
        }
        + T_{e\mathrm{,t}}^{7/2}
    \right)^{2/7},
\end{equation} we obtain a good match to the simulation results as shown in figure \ref{fig:Teomp_breakdown}.b (labeled \enquote{Match to sim.}), where
\begin{equation}
    \label{eq:placeholder_label}
    \langle q_{e\parallel \text{cond}} \rangle \equiv \frac{1}{L_\parallel} \int_{\text{OMP}}^{\text{OT}} q_{e\parallel \text{cond}} \, \mathrm{d}s_\parallel 
\end{equation} is the parallel average of the parallel electron conducted heat flux density along the peak SOL flux tube, and 

\begin{equation}
    \label{eq:placeholder_label}
    \left\langle \kappa_{e\parallel 0 \mathrm{,FL}} \right\rangle
    \equiv
    \frac{
        \int_{\mathrm{OMP}}^{\mathrm{OT}} \kappa_{e\parallel 0 \mathrm{,FL}} q_{e\parallel \mathrm{cond}} \, \mathrm{d}s_{\parallel}
    }{
        \int_{\mathrm{OMP}}^{\mathrm{OT}} q_{e\parallel \mathrm{cond}} \, \mathrm{d}s_{\parallel}
    }
\end{equation} is the parallel average of the parallel electron heat conductivity divided by $T_{e}^{5/2}$, which includes the flux limiting corrections. The flux limiting procedure in SOLPS prevents unrealistically high heat fluxes at low collisionalities by adding a correction factor to the classical (or Spitzer-H{\"a}rm) heat flux density, such that 
$q_{e\parallel \text{cond}} = - \kappa_{e \parallel 0 \mathrm{,FL}} T_{e}^{5/2}\nabla_{\parallel} T_e$, where $\kappa_{e \parallel 0 \mathrm{,FL}}=c_{\mathrm{FL}}\kappa_{e \parallel 0 \mathrm{,classical}}$ and $c_{\mathrm{FL}}$ contains the flux limiting model (for more details see \cite{day1996effect,stangeby2000plasma,fundamenski2005parallel}).   

The \Teomp{} trends in figure \ref{fig:Teomp_breakdown}.a and b can be divided into three regimes: (i) \Tet{} $>$ 10 eV: the flux limited regime, (ii) 1$<$ \Tet{}$<$ 10 eV: the partially detached regime and (iii) \Tet{}$<$ 1 eV: the pronounced detachment regime. Overall, \Teomp{} is moderated by the \kappaeFL{} dependence on \Tet, as well as decreasing \qparavg{} with decreasing \Tet{} through a combination of variation in the convective to conductive \Psep{} ratio, and through dissipation in the SOL-divertor. The $T_{e\mathrm{,t}}^{7/2}$ term in equation \ref{eq:Tesep_sim_match} has a negligible effect for the \Tet{} range considered and can be neglected. In the flux limited regime at the highest \Tet, \kappaeFL{} $\approx 1300 \: \mathrm{Wm^{-1}eV^{-7/2}}$, rising to $2300 \: \mathrm{Wm^{-1}eV^{-7/2}}$ as \Tet{} is reduced from 70 to 3 eV, which is close to the typically assumed constant value of $2600 \: \mathrm{Wm^{-1}eV^{-7/2}}$ for unseeded conditions. Neglecting the flux limits in $\left\langle \kappa_{e\parallel0} \right\rangle$, as is typically done in estimating \Teomp{} using the Spitzer-H{\"a}rm power balance approach in experiments (e.g., \cite{stangeby2000plasma, stangeby2015identifying,eich2021separatrix,silvagni2025separatrix}), has a significant impact as shown in figure \ref{fig:Teomp_breakdown}.b comparing the \enquote*{Match to sim} and \enquote*{$\left\langle \kappa_{e\parallel0} \right\rangle$=2300} lines. For \Tet{}$>$ 10 eV, replacing \qparavg{} with the total plasma heat flux density $\langle Q_{\parallel \text{plasma}} \rangle$ leads to an increase in \Teomp{} compared to the SOLPS values, whereas in the partially detached regime \qparavg$\approx\langle Q_{\parallel\text{plasma}} \rangle$ appears to be valid, indicating the dependence of the conducted parallel electron heat flux fraction, $f_{\mathrm{cond}}$, on \Tet.  

 In figure \ref{fig:Teomp_breakdown}.b the differences in the \Teomp{} trends are also shown considering different fueling locations; namely core fueling vs fueling from the inner SOL (see figure \ref{fig:solps_grid} for the fueling locations). In the partially detached regime, the fueling location change from core to inner SOL leads to a more moderate \Teomp{} reduction with decreasing \Tet. The steeper \Teomp{} reduction in the core fueling cases arises due to the significant core particle fluxes required to achieve higher densities. This leads to a large convective \Psep{} component, 

\begin{equation}
    \label{eq:placeholder_label}
    P_{\text{sep,LFS,conv}} \approx \frac{5}{2}\,\Gamma_{\text{D,core}}\,T_{e,\text{OMP}}\left(1 + \frac{T_{i,\text{OMP}}}{T_{e,\text{OMP}}}\right),
\end{equation}
and hence reduces the conducted \Psep{} component (and \Teomp{} through eq. \ref{eq:Tesep_sim_match}) for a constant total \Psep. The increasing core particle flux also drives down \Teomp{} as it increases faster than $P_{\text{sep,LFS,conv}}$ since as $\Gamma_{\text{D,core}}$ increases, \Teomp{} must decrease, assuming constant $T_{i\text{,OMP}}$/\Teomp. Below \Tet{}$=1$ eV, \Teomp{} decreases further due to dissipation and thus continued reduction in \qparavg{} at a relatively constant \kappaeFL. 

Having identified the main drivers leading to a significant decrease in \Teomp{} with increasing \neomp{} (and decreasing \Tet) through equation \ref{eq:Tesep_sim_match}, it's important to consider to what extent the  Spitzer-H{\"a}rm (S-H) power balance approach used to infer \Tesep{} in experiments captures the above trends. The extrapolated model for the SPARC SepOS operational boundaries shown in figure \ref{fig:sepos_12Tvs8T_rollover} was developed using a combination of AUG experimental datasets and heuristic techniques underpinned by the drift-Alfv\'en (DALF) model \cite{eich2021separatrix,eich2025separatrix}. Since the S-H power balance approach was extensively used for evaluating \Tesep{} and locating the separatrix in these experiments from kinetic electron temperature and density profile measurements in the plasma edge, the methodology and the inferred \Tesep{} and \nesep{} estimates thus directly inform the SepOS formulation.

Using the S-H power balance model following \cite{eich2020turbulence,eich2021separatrix}, we next evaluate $T_{\text{e},\text{sep}}^{\text{SH}}\approx T_{\text{e},\text{OMP}}^{\text{SH}}$ from

\begin{equation}
    T_{\text{e},\text{OMP}}^{\text{SH}} \approx 0.85 \left[ \frac{7 f_{\text{out}} f_{\text{cond}} f_{e} P_{\text{sep}} \hat{q}_{\text{cyl}}^2 A}{8 \kappa_{e \parallel \text{0}}  \hat{\kappa} \lambda_{\text{q}}^{\text{OMP}}} \right]^{\frac{2}{7}}
    \label{eq:Teomp_SH},
\end{equation} where $\hat{\kappa} = \sqrt{\frac{1 + \kappa_{\text{geo}}^{2}\left(1 + 2\delta^{2} - 1.2\delta^{3}\right)}{2}}$, $A=R/a$ is the aspect ratio, $f_{\text{out}}=0.6$, $f_{\text{cond}}=0.7$ and $f_{e}=0.8$ are defined as the fraction of the total power entering the SOL that is directed towards the outer divertor, the fraction of conducted power and the fraction of power transported by electrons, respectively. The power fraction values were selected based on typically used values in experiments (e.g,. \cite{silvagni2025separatrix}), and slightly modified together with the 0.85 pre-factor to obtain a better match to the SOLPS output values in figure \ref{fig:Teomp_breakdown}.b. $\lambda_{\text{q}}^{\text{OMP}}=\left( \frac{9}{16} \right) \langle \lambda_{\text{q}} \rangle$ is the local power decay width at the OMP using the Eich scaling from \cite{eich2020turbulence}, given as a poloidally averaged quantity approximated by $\langle \lambda_{\text{q}} \rangle \approx 2/7 \langle \lambda_{\text{T}_{e}} \rangle$
with 

\begin{equation}
    \langle \lambda_{\text{T}_e} \rangle (\mathrm{m}) =  2.1  \left( 1 + 2.1 \alpha_{t}^{1.7} \right) \rho_{s,\mathrm{pol}}
    \label{eq:lambda_Te}, 
\end{equation} where the poloidal ion sound Larmor gyro-radius given by

\begin{equation}
    \rho_{s,\mathrm{pol}} = \frac{\sqrt{m_{\mathrm{D}} T_{e\mathrm{,OMP}}}}{e B_{\mathrm{pol}}}
    \label{eq:pol_rho},
\end{equation}$B_{\text{pol}}$ is the poloidally averaged poloidal magnetic field at the separatrix and $m_{\text{D}}$ is the deuterium ion mass. A constant parallel electron heat conduction coefficient is used with a finite \Zeff{} correction, 
\begin{equation}
    {\kappa_{e,0}}\approx 2600/f_{\kappa,0}(Z_{\text{eff}})
    \label{eq:kappae0_simple}
\end{equation}  with $f_{\kappa,0}(Z_{\text{eff}})=0.672+0.076 Z_{\text{eff}}^{0.5}+0.252Z_{\text{eff}}$ \cite{eich2021separatrix,goldston2017new}. For the unseeded density scans, we assume a constant \Zeff=1.25 for consistency with the AUG SepOS datasets\footnote{One minor difference is that in the SepOS formulation in \cite{eich2021separatrix} the reference point for \Tesep{} and \nesep{} is taken slightly inside the separatrix at $\rho_{\text{pol}}=0.999$, whereas we evaluate \Teomp{} and \neomp{} at the peak flux tube just outside the separatrix.}. 

An important distinction in the S-H model used in \cite{eich2020turbulence,eich2021separatrix} is the assumption on how the power enters the SOL, as recently clarified in Appendix A of \cite{eich2026ARC}, and originally described in Chapter 5 of \cite{stangeby2000plasma}. To obtain a better match to the SOLPS \Teomp{} results we assume the 7/2 factor (in eq. \ref{eq:Teomp_SH} this leads to the 7/8 factor) rather than 7/4 used in \cite{eich2020turbulence,eich2021separatrix}, where in the latter the power is assumed to enter the SOL along its entire length rather than at the OMP. The distinction is somewhat academic given the approximate nature of the basic 2PM, but it has a potentially important consequence on interpreting the separatrix location in experiment and estimating \alphat{} values with the inferred separatrix plasma parameters. Using the 7/4 factor (i.e., 7/16 in eq. \ref{eq:Teomp_SH}) would  underestimate \Teomp{} in figure \ref{fig:Teomp_breakdown}.b relative to the SOLPS values, which would then also underestimate \neomp{} based on the inferred position of the separatrix further outboard. In Appendix A of \cite{eich2026ARC} this leads to a correction of the \alphat{} QCE empirical threshold value from 0.55 to 0.43, which would result in a moderate shift of the \alphat{} contour in figure \ref{fig:sepos_12Tvs8T_rollover} towards lower densities. Although we continue to use the \alphat{}$=$ 0.55 QCE threshold throughout the paper, we note that in light of the nuanced \TeompSH{} interpretation this is perhaps a conservative assumption.   

We use an iterative procedure to estimate \TeompSH{} and $\alpha_{\text{t}}$ that does not make any assumptions on the position of the separatrix in the SOLPS edge profiles, and instead uses an initial \TeompSH{} guess of 100 eV to locate the separatrix, and the accompanying \neompSH{} value at the same position. \TeompSH{} is then iteratively updated using equations \ref{eq:Teomp_SH}-\ref{eq:pol_rho}, thus following a similar approach to interpreting edge profiles in experiment. This procedure yields the \TeompSH{} estimates labeled as \enquote*{S-H power balance w/ $\lambda_{\text{q}}(\alpha_{\text{t}})$}  shown in figure \ref{fig:Teomp_breakdown}.b. The S-H estimates are in relatively good agreement, albeit slightly below, the \enquote*{$\left\langle \kappa_{e\parallel0} \right\rangle$=2300} estimates, and also reproduce the reduction in \Teomp{} below \Tet{} of 1 eV.  This reduction can be attributed to increasing $\alpha_{\text{t}}$ in equation \ref{eq:lambda_Te} and thus an effective widening of \lamq. The generally good agreement is perhaps not surprising, since we selected \chiperp{} values following the empirical AUG \lamq{} scalings from \cite{eich2020turbulence}, aiming for \lamq$\approx$0.3-0.6 mm for the SPARC 12.2 T scenario. However, as the comparison in figure \ref{fig:Teomp_breakdown}.b shows, neglecting the impact of the flux limits on $\kappa_{e\parallel0}$ could lead to moderate \TeompSH{} underestimates (and by extension uncertainty in the separatrix location and \neompSH). Since the discrepancy is most pronounced at low edge collisionalities or high \Tet{} above tens of eV (up to about 20\% lower \TeompSH{} compared to the SOLPS values), we don't expect a significant impact on the SepOS formulation. This could in principle be further quantified by applying a SOLPS-derived $\kappa_{e\parallel0,\text{FL}}$ dependence on \Tet{} or \alphat{} in interpreting the experimental data on existing machines, within the limits of the accuracy of the flux limiting approach used in SOLPS. 

The evolution  of the \qparavg{} profile, mapped to the outer mid-plane, with increasing \alphat{} is shown in figure \ref{fig:Teomp_breakdown}.c. Although a full energy balance is outside the scope of this contribution, it's clear from the reduction and flattening of the \qparavg{} radial profile with increasing \alphat{} that the corresponding \lamq{} widening is independent of any radial turbulent transport considerations, since the SOLPS scans are carried out at fixed \chiperp{} and \Dperp. It's important to note that simulations with drifts activated are needed to further evaluate the net impact on \lamq{} at fixed \chiperp{} and \Dperp. Nevertheless, in the present analysis, we conclude that the observed widening of \lamq{} at fixed \chiperp{} and \Dperp{} due to a reduction in \qparavg{} via convective effects and dissipation in the SOLPS scans can be correlated with the experimentally observed widening in AUG \cite{eich2020turbulence,faitsch2023analysis}, which, in experiment, has been attributed to increased turbulence near the separatrix. Disentangling this interplay of parallel transport, anomalous cross-field transport and downstream dissipation mechanisms requires further interpretive studies on existing devices.  

Returning now to the correlation between \Tet{} and \alphat{} $\propto n_{e\mathrm{,OMP}}/T_{e\mathrm{,OMP}}^2$ shown in figure \ref{fig:Tet_alphat}.a, we combine equation \ref{eq:neomp} with the general \Teomp{} $\propto q_{\parallel\text{,u}}^{2/7} L_{\parallel}^{2/7}$ scaling according to S-H parallel electron conduction model to obtain 

\begin{equation}
    \alpha_{\mathrm{t}}  \propto  T_{e\mathrm{,t}}^{-1/2}q_{\parallel ,u}^{1/7} L_{\parallel}^{-6/7}  \frac{(1-f_{\mathrm{pwr}})}{(1-f_{\mathrm{mom}})}.
    \label{eq:alphat_Tet}
\end{equation}We can neglect the weak $q_{\parallel\text{,u}}^{1/7}$ dependence as well as the $L_{\parallel}^{-6/7}$ dependence for fixed \qcyl, focusing on the remaining strong \fpwr{} and \fmom{} dependence on \Tet{} as well as the explicit $T_{e\mathrm{,t}}^{-1/2}$ dependence. The result is the line labeled \enquote{Basic 2PM and S-H scaling (eq. \ref{eq:alphat_Tet})} shown in figure \ref{fig:Tet_alphat}.b. The monotonically increasing trend with decreasing \Tet{} approximates the \alphat{} dependence, but it does not capture the more rapid \alphat{} rise at lower \Tet{} values. Hence, while the basic scaling from eq. \ref{eq:alphat_Tet} is an approximation, it is informative as it links \alphat{} directly to the plasma-neutral dissipation close to the target captured by the \fpwr{} and \fmom{} losses shown in figure \ref{fig:fmomfpwr}, and without explicit \Pin{} and \chiperp{} normalizations since \qpar{} largely cancels. If instead we combine eq. \ref{eq:neomp} with the \enquote{Match to sim.} \Teomp{} estimates from eq. \ref{eq:Tesep_sim_match} (figure \ref{fig:Teomp_breakdown}.b) and include the \Pin$^{-0.36}$ and \chiperp$^{0.25}$ normalization, we obtain the line labeled \enquote{\neomp{} eq. \ref{eq:neomp}; \Teomp{} eq. \ref{eq:Tesep_sim_match}}. This yields a better approximation to the \alphat{} trend since it also captures the observed decrease in \Teomp{} with decreasing \Tet, as discussed above. 

The flexibility in the choice of mapping parameter (either \Tet{} or \alphat) is therefore quite convenient for connecting the QCE access condition with power exhaust constraints. For example, for the divertor quantities \Idiv{} and \qdep, the \Tet{} mapping enables directly capturing additional constraints like the effective sputtering yield estimates for W PFCs, which are a strong function of \Tet  \cite{eckstein2002calculated,stangeby2018basic}. On the other hand, recasting the dissipative divertor properties driven by atomic/molecular physics loss channels that scale with $n_{e}$ and $T_{e}$ to \alphat{} (or edge collisionality \nuedge{} since \alphat{} $\propto$ \qcyl\nuedge), facilitates a simple description for linking the power exhaust constraints to the turbulent transport considerations at the separatrix and the observed near-SOL decay width broadening dependence on \alphat{} or \nuedge{} \cite{eich2020turbulence,faitsch2023analysis}. However, since \qcyl{} is fixed in our analysis, it's not clear how changes in shaping parameters will impact the \alphat{} and \Tet{} correlation beyond the expected $L_{\parallel}^{-6/7}$ dependence in eq. \ref{eq:alphat_Tet} for a given $B_{\text{t}}$ and $I_{\text{p}}$. Therefore, the foregoing analysis only establishes the quantitative link between \Tet{} and $n_{e\mathrm{,OMP}}/T_{e\mathrm{,OMP}}^2$. 

Varying the \Dperp/\chiperp{} ratio is likely to result in a vertical shift of the trends in figure \ref{fig:Tet_alphat} since such variation will also impact $n_{e\mathrm{,OMP}}/T_{e\mathrm{,OMP}}^2$ and the ratio of the near-SOL density and temperature gradient lengths, $\eta_{e}=\lambda_{n_{e}}/\lambda_{T_{e}}$. In AUG $\eta_{e}$ was found to be close to unity over a large range of plasma parameters, although higher $\eta_{e}$ was observed in detached conditions \cite{sun2020empirical}. This was associated with dominant convective filamentary transport at high densities resulting in flatter density profiles, and decoupling of the cross-field particle and heat transport mechanisms. At fixed \Dperp/\chiperp{} these effects are not captured in the modeling. Hence we can expect a moderate modification to the \alphat{} vs \Tet{} trend at \Tet{}$<$ 1-2 eV around the \Idiv{} rollover corresponding to particle detachment, but it is unclear whether this would result in de-correlation of the trends over the \Pin{} and \chiperp{} range, or whether the $\eta_{e}$ increase with detachment is also correlated to \Tet. 

\section{Extending analysis to Ne seeding}

\begin{figure}
    \centering
    \includegraphics[width=.95\linewidth]{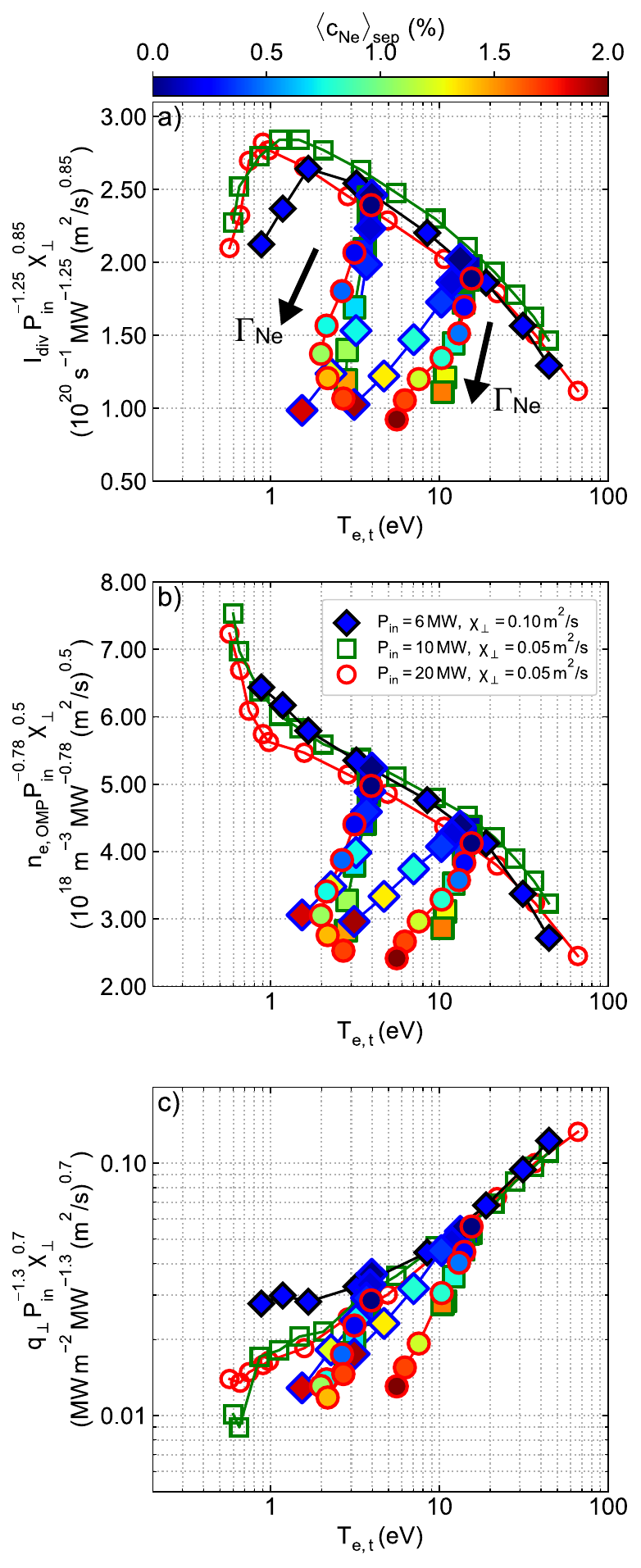}
    \caption{Dependence of a-c) normalized \Idiv, \neomp{} and \qdep{} on \Tet{} for select unseeded density scans and Ne seeding scans at fixed \GamD. The \enquote*{low recycling} Ne seeding branch starts at the corresponding \Tet{}$\approx$ 15 eV unseeded cases, whereas the \enquote*{high recycling} Ne seeding branch starts at the corresponding \Tet{}$\approx$ 4 eV unseeded cases, with the marker fill color of the Ne branches indicating \cNe.}
    \label{fig:Idiv_neomp_Tet_Ne}
\end{figure}

In extending a subset of the pure deuterium density scans with Ne seeding, we take advantage of the self-similar \Tet{} trends for the unseeded density scans, and introduce a \GamNe{} scan while maintaining constant \GamD{} corresponding to a particular unseeded reference value of \Tet. In this way, we can select the same \Tet{} point for the \Pin{} and \chiperp{} normalized parameters of interest such as \neomp, \Idiv{} and \qdep, and map out the Ne seeding scans or \enquote{branches}, as shown in figure \ref{fig:Idiv_neomp_Tet_Ne}. We refer to the Ne seeding branch starting at about \Tet$=$15 eV as the low-recycling Ne branch, and the branch at \Tet$=$4 eV as the high-recycling Ne branch for convenience. Using the derived \Pin{} and \chiperp{} normalizations, we again observe self-similar trends extending beyond the unseeded density scans to the Ne seeding branches, with approximate correspondence between \cNe, and \Tet{} over the \Pin{} and \chiperp{} range. This correspondence is notably more exact for the \Pin{}= 10 MW and 20 MW cases with \chiperp{}= 0.05 $\mathrm{m^2/s}$, whereas the \Pin{} = 6 MW cases with \chiperp{}= 0.1 $\mathrm{m^2/s}$ exhibit relatively more deviation, especially for the low-recycling Ne branch, suggesting a modest transport dependence not fully captured by the derived \chiperp{} normalization.   

\begin{figure*}[t]
    \centering
    \includegraphics[width=0.8\linewidth]{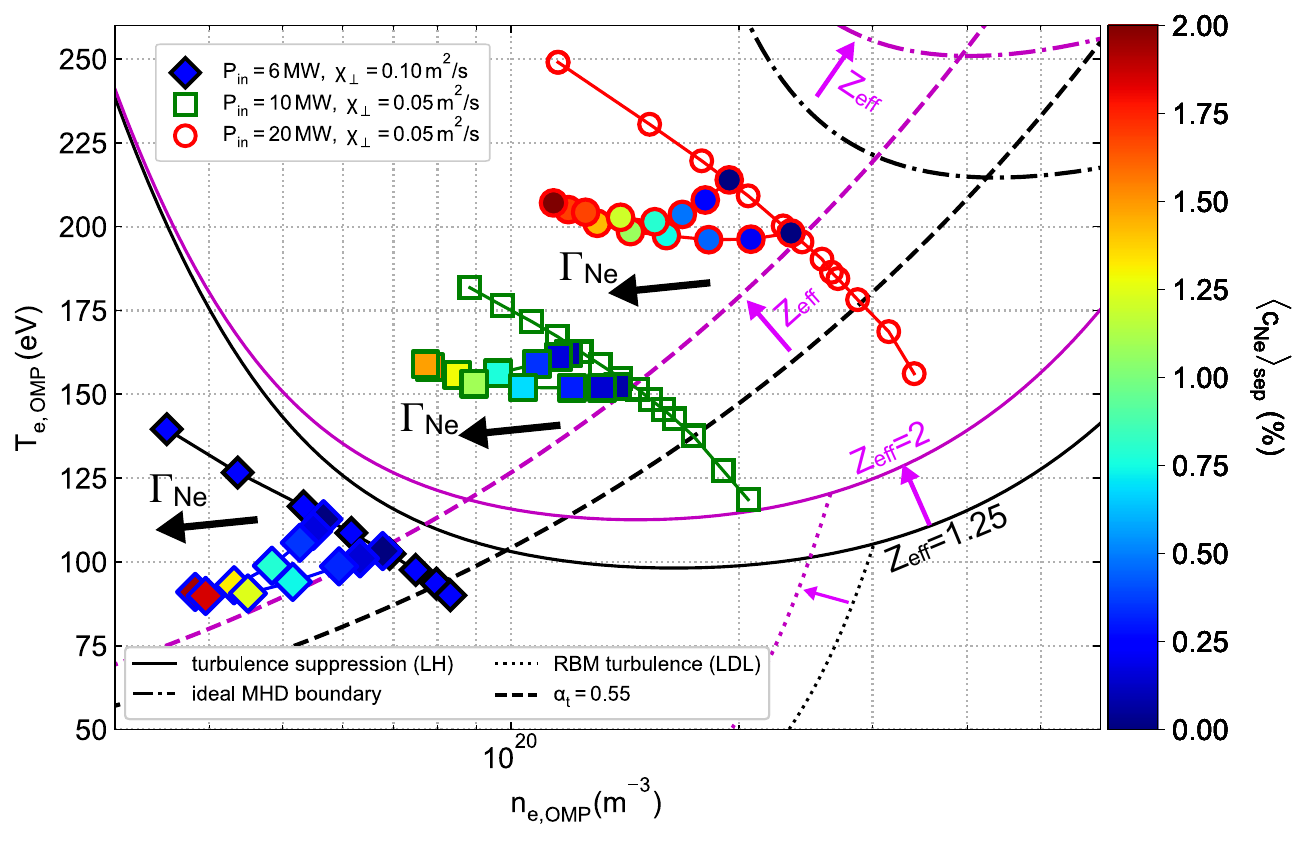}
    \caption{The projected SPARC separatrix operational space boundaries and \alphat{}$=$ 0.55 contours calculated assuming \Zeff{}$=$ 1.25 (black) and \Zeff{}$=$ 2 (magenta) for the full field $B_{\text{t}}=$12.2 T, $I_{\text{p}}=$8.7 MA scenario. Markers correspond to select unseeded density scans and the \enquote*{low recycling} and \enquote*{high recycling} Ne seeding branches at different \Pin{} and \chiperp{} values. The marker fill color of the Ne branches indicates \cNe.}
    \label{fig:sepos_Ne}
\end{figure*}

As shown in figure \ref{fig:Idiv_neomp_Tet_Ne}, increasing \GamNe{} (and hence \Zeff{} and \cNe{} up to 2\%) leads to a pronounced reduction in \neomp, \Idiv{} and \qdep, and a corresponding moderate reduction in \Tet. The \neomp{} reduction with increasing \GamNe{} can be attributed to the so called \enquote*{power limitation} or \enquote*{power starvation} effect. Qualitatively, increasing \cNe{} increases impurity radiation in the divertor region, limiting the available power for neutral molecular dissociation and ionization/excitation, leading to a reduction in the recycling fluxes at the target with knock-on effects on reducing the upstream plasma pressure. Similar observations were made in the SOLPS modeling studies underpinning the ITER tungsten divertor physics basis (\cite{pitts2019physics}, and references therein). A straightforward, but incomplete, interpretation of this effect is obtained through simple power balance considerations. Assuming that the change in \Tet{} in the \GamNe{} scan is small (i.e., an assumption which allows us to disregard the role of momentum losses), and that at high \cNe{} the Ne radiation power losses dominate, \fpwr$\approx(1-f_{\text{rad,Ne}})\propto (T_{e\mathrm{,t}}\Gamma_{\mathrm{\parallel,t}})$/\qpar. To satisfy the power balance and assuming a relatively constant \qpar{} (valid in the present SOLPS datasets), the reduction in $(1-f_{\text{rad,Ne}})$ is balanced by a reduction in the recycling flux and thus \qdep{} and plasma pressure at the target.  The upstream pressure is then reduced through parallel pressure balance for a given \Tet{} point on the \fmom{} trend in figure \ref{fig:fmomfpwr}.a. We leave the more detailed explanation combining power and momentum balance as an exercise for future work, focusing instead on the impact of Ne seeding on the SepOS.

\subsection{Implications of separatrix density reduction with Ne seeding}
In the following, we demonstrate how the reduction in \neomp{} with increasing \cNe{} opens up another optimization parameter in the PE-SepOS framework. The impact on the SepOS is shown in figure \ref{fig:sepos_Ne}, where the Ne seeding branches have been added to the subset of unseeded density scans shown in figure \ref{fig:sepos_12Tvs8T_rollover}. The impact of \Zeff{} on the SepOS boundaries is also shown, with a shift in the \alphat{} $=$ 0.55 empirical QCE threshold contour towards lower densities as \Zeff{} is increased from 1.25 to 2\footnote{Note that the unseeded SepOS boundaries have been calculated with \Zeff{}$=$ 1.25 for consistency with the SepOS experimental datasets from AUG \cite{eich2021separatrix}, where, in contrast to the SOLPS unseeded density scans in a pure D plasma, such pure plasmas are difficult to achieve experimentally.}. We note that the \Zeff{} impact on the LH, LDL, and ideal MHD boundaries as given by the SepOS framework has not been experimentally tested. Furthermore, the impact of Ne seeding on the core radiation inside of the SOLPS computational grid core boundary is neglected, which can also influence H-mode access. 

\begin{figure}
    \centering
    \includegraphics[width=0.9\linewidth]{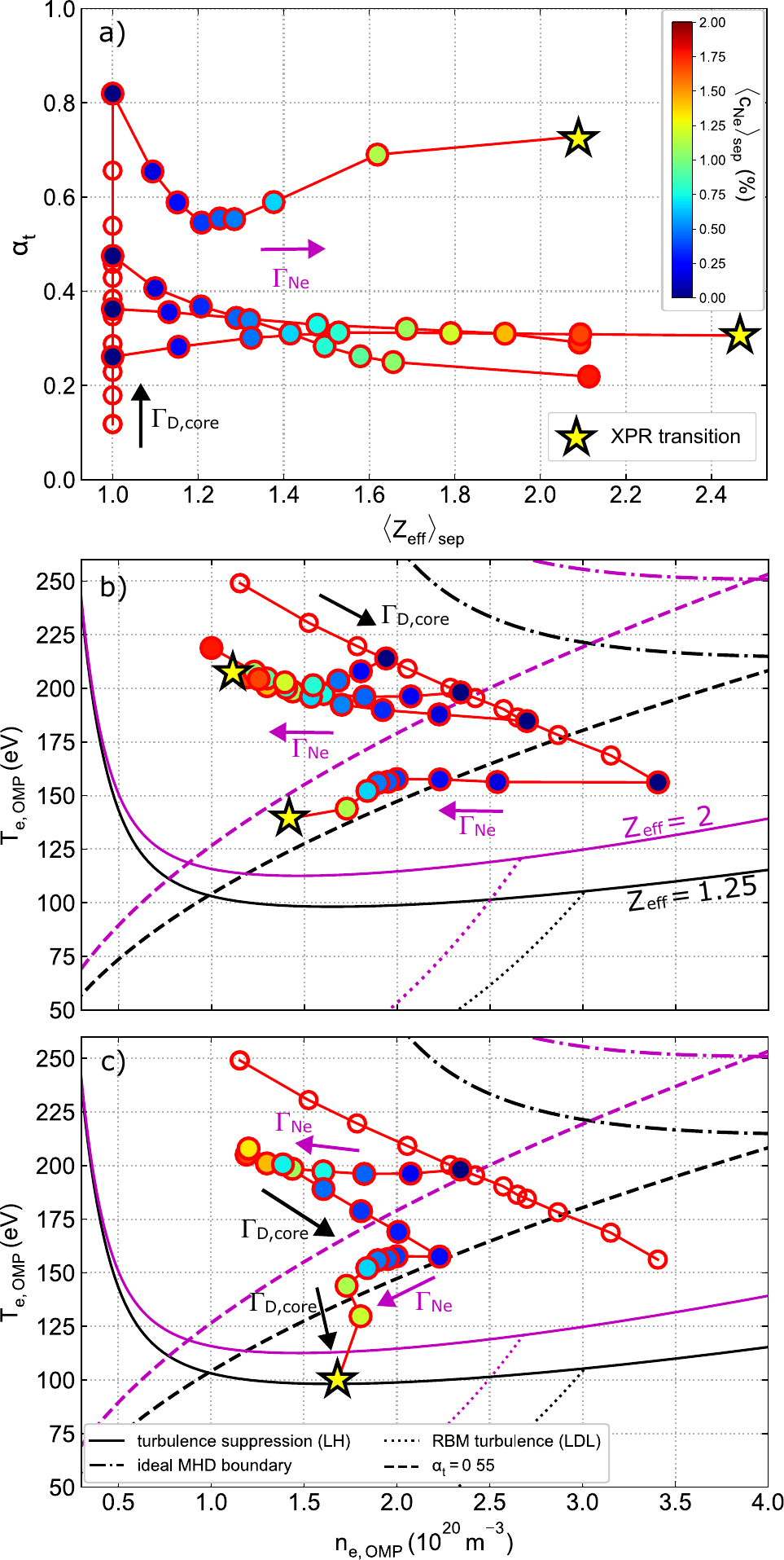}
    \caption{a) Dependence of \alphat{} on \Zeff{} and \cNe{} for the different \Pin{}$=$ 20 MW, \chiperp{}$=$ 0.05 $\mathrm{m^2/s}$ Ne seeding branches; b) projected SepOS for the same Ne seeding branches and unseeded density scan; c) density recovery from the \enquote*{low recycling} Ne seeding branch through additional \GamD, and subsequent further \GamNe{} seeding and \GamD{} fueling until XPR transition near the LH boundary.  Empty markers correspond to the unseeded \Pin{}$=$ 20 MW, \chiperp{}$=$ 0.05 $\mathrm{m^2/s}$ density scan. Marker fill colors indicate \cNe. \enquote*{XPR transition} cases correspond to cases with pronounced radiation inside the confined plasma. The projected SepOS boundaries and \alphat{}$=$ 0.55 contours in b) and c) are calculated assuming \Zeff{}$=$ 1.25 (black) and \Zeff{}$=$ 2 (magenta) for the full field $B_{\text{t}}=$12.2 T, $I_{\text{p}}=$8.7 MA scenario.}
    \label{fig:alphat_Zeff}
\end{figure}

The common feature for all Ne seeding branches is a relatively small impact on \Teomp{} (decreasing or increasing depending on \cNe{} and also \chiperp) for a large corresponding monotonic reduction in \nesep. Figure \ref{fig:alphat_Zeff}.a shows an example of the impact of the Ne seeding branches on \alphat$\propto (n_{e\mathrm{,OMP}}Z_{\text{eff,OMP}})/T_{e\mathrm{,OMP}}^2$, where two additional Ne seeding branches have also been added at higher densities (with \alphat{}$>$ 0.4 and \neomp{}$>$ $2.5\times10^{20}\:\mathrm{m^{-3}}$) for the \Pin{}$=$ 20 MW, \chiperp{}$=$ 0.05 $\mathrm{m^2/s}$ scan to further evaluate the \Zeff{} dependence.  The \neomp{} reduction is to some extent balanced by the increase in \Zeff, and to a lesser extent by the more modest changes in \Teomp. The decrease and subsequent increase in \alphat{} with increasing \Zeff{} is particularly significant for the highest density Ne seeding branch, which could potentially enable high \alphat{} access at high \cNe, although the trends shown in figure \ref{fig:alphat_Zeff}.a need to be further refined through drift-activated simulations to evaluate the possible Ne redistribution impact. 

At sufficiently high \cNe{} and \Zeff{} $>$ 2.0, the Ne radiation front incursion towards the X-point along the inner and outer divertor legs can transition to an X-point radiator (XPR) plasma solution with the peak radiation localized to the confined plasma region inside the X-point. These transitions are marked in figure \ref{fig:alphat_Zeff}, where the higher density Ne seeding branches are also shown relative to the \alphat{}$=$ 0.55 QCE threshold. The proximity of the XPR transition to the LH boundary in the higher density Ne seeding branch is qualitatively consistent with XPR experiments on JET and AUG, which can exhibit H-L back transitions depending on the impurity species mix \cite{bernert2017power}, or ELM suppression exhibiting a weak pedestal with remaining lack of clarity on the confinement regime \cite{bernert2021x}.

The consequence of a decreasing or roughly constant \alphat{} along the Ne seeding branches could present challenges for QCE/EDA H-mode access, despite a concomitant favorable shift in the \alphat{}$=$ 0.55 contour to lower densities with increasing \cNe{} or \Zeff{} in figure \ref{fig:sepos_Ne}. Accessing dissipative regimes via Ne seeding thus leads to more modest \alphat{} values relative to the unseeded cases for which detachment is achieved through plasma-neutral interactions at higher \GamD, and therefore \neomp{} and \alphat. In Section \ref{sec:lowhighpwrproj} we consider these tradeoffs in evaluating the accessible operational points with respect to QCE, heat flux mitigation and sputtering suppression at the divertor targets. 

The \neomp{} reduction at high \cNe{} can be recovered to some extent with additional fueling for fixed \GamNe, as shown in figure \ref{fig:alphat_Zeff}.c. Doing so increases the level of detachment on the outer target, and raises \neomp{} closer to the high \alphat{} values required for QCE access, while also reducing \cNe{} and \Zeff. However, this comes at the cost of further \Teomp{} reduction, thus placing the operational points closer to the LH boundary. With further fueling and Ne seeding, the pronounced level of detachment can lead to a transition to the XPR regime as shown in figure \ref{fig:alphat_Zeff}.c, consistent with experience from experiments. 

\begin{figure}
    \centering
    \includegraphics[width=0.85\linewidth]{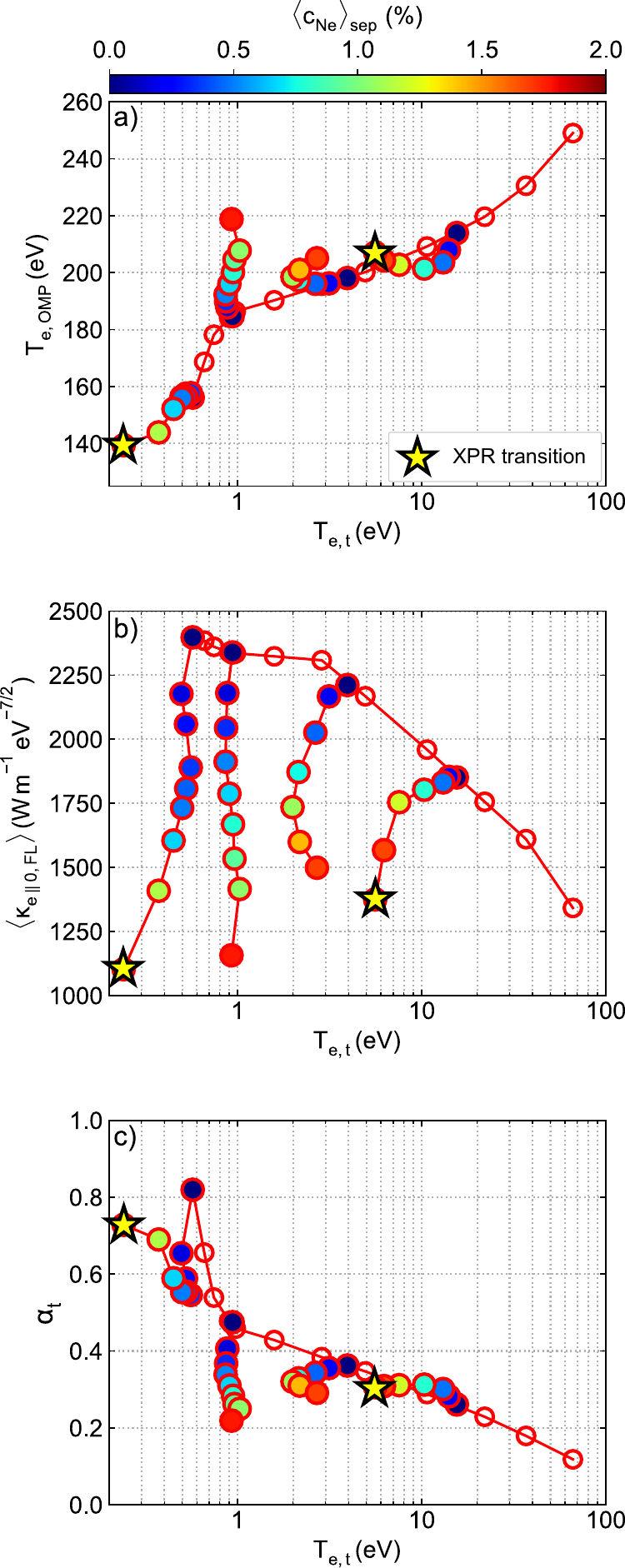}
    \caption{Dependence of a-c) \Teomp, \kappaeFL{} and \alphat{} on \Tet{} for the density scan and Ne seeding branches at \Pin{}$=$ 20 MW, \chiperp{}$=$ 0.05 $\mathrm{m^2/s}$. Empty markers correspond to the unseeded cases whereas marker fill colors indicate \cNe. \enquote*{XPR transition} cases correspond to cases with pronounced radiation inside the confined plasma.}
    \label{fig:Teomp_Ne_branches}
\end{figure}

\subsection{Impact of Ne seeding on \Teomp}
The \Teomp{} trends along the Ne seeding branches shown in figures \ref{fig:sepos_Ne} and \ref{fig:alphat_Zeff}, and hence the impact on \alphat$\propto T_{e,\text{OMP}}^{-2}$, can be explained in the context of eq. \ref{eq:Tesep_sim_match} through a combination of i) the finite \Zeff{} correction to $\kappa_{e \parallel 0}$  (eq. \ref{eq:kappae0_simple} and \kappaeFL{} in  eq. \ref{eq:Tesep_sim_match}), ii) up to 20\% reduction in \Psep{} at \Zeff{}$=$ 2 with Ne seeding; and iii) radiative dissipation along the peak flux tube, with both ii) and iii) contributing to a reduction in \qparavg. The impact of Ne seeding on the core radiated power inside the computational grid boundary is neglected. The recovered \Teomp{} trends for the Ne seeding branches using eq. \ref{eq:Tesep_sim_match} are shown in figure \ref{fig:Teomp_Ne_branches}.a with exact agreement to the SOLPS values. The finite \Zeff{} correction used in SOLPS according to the Zhdanov formulation \cite{coster_solps_manual} is similar to eq. \ref{eq:kappae0_simple}, with the calculated \kappaeFL{} from SOLPS output shown in figure \ref{fig:Teomp_Ne_branches}.b. The finite \Zeff{} correction to \kappaeFL{} largely balances the reduction in \qparavg{} for the two Ne seeding branches at higher \Tet. For the Ne seeding branch at \Tet{}$=$ 1 eV the reduction in \kappaeFL{} evidently dominates over the reduction in \qparavg, whereas the opposite is true for the Ne seeding branch at the lowest \Tet.  The overall impact on \alphat$\propto (n_{e\mathrm{,OMP}}Z_{\text{eff,OMP}})/T_{e\mathrm{,OMP}}^2$ is shown in figure \ref{fig:Teomp_Ne_branches}.c. The \alphat{} vs \Tet{} correlation obtained for the unseeded density scans is now broken, which we attribute to the combination of the Ne seeding impacts on \Teomp{} as above, in addition to the pronounced reduction in \neomp. This is consistent with both the finite \Zeff{} correction and the volumetric Ne radiation being more decoupled from \Tet{} in contrast to the neutral-plasma interaction momentum and power loss channels close to the divertor target.

\section{\label{sec:lowhighpwrproj}Projecting self-similar operational points from low to high input power}
\begin{figure}[t]
    \centering
    \includegraphics[width=0.75\linewidth]{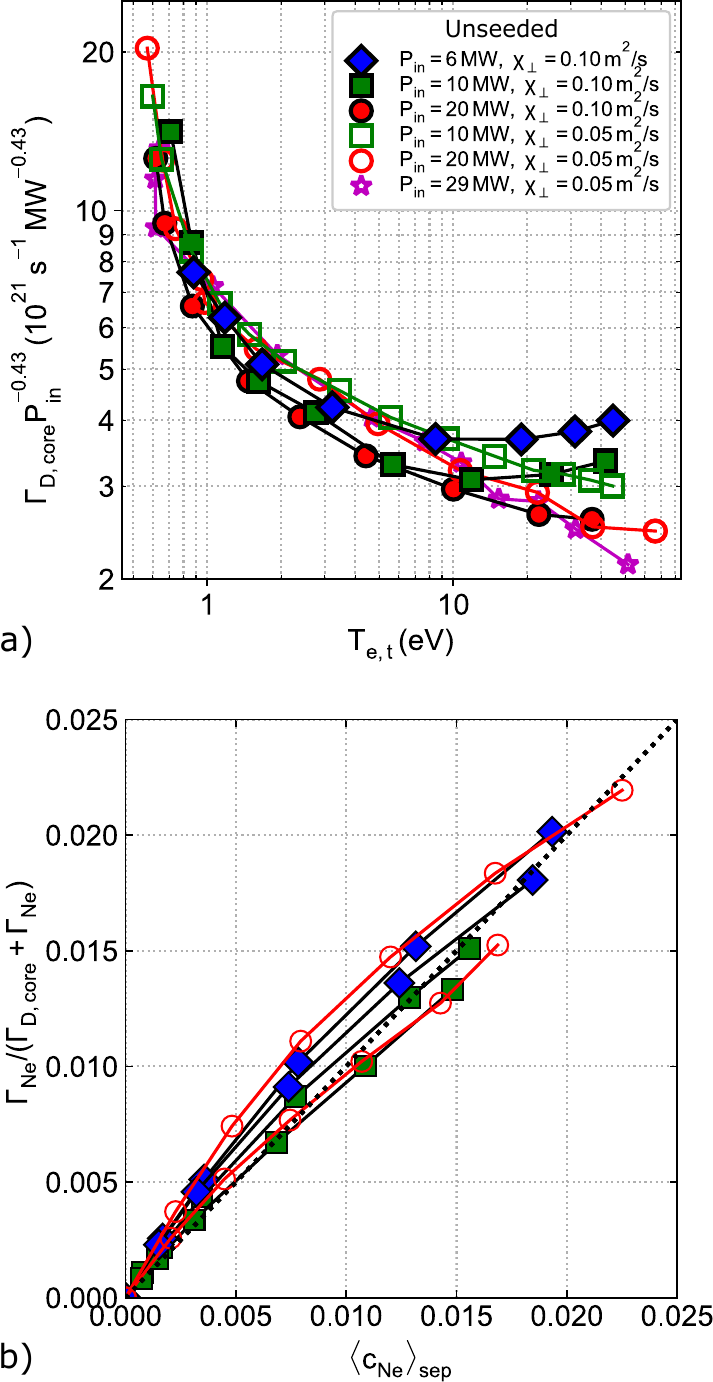}
    \caption{Dependence of a) \GamD$P_{\text{in}}^{-0.43}$ on \Tet{} for the unseeded density scans at different \Pin{} and \chiperp{} values; b) Dependence on the \GamD/(\GamD{} + \GamNe) ratio on \Tet{} for the \enquote*{low recycling} and \enquote*{high recycling} Ne seeding branches shown in figure \ref{fig:sepos_Ne}.}
    \label{fig:phys_params_to_actuators}
\end{figure}

The self-similar trends mapped to \Tet{} in figures  \ref{fig:Idiv_neomp_Tet}, \ref{fig:fmomfpwr} and \ref{fig:Idiv_neomp_Tet_Ne} can potentially be utilized to project the accessible operational points from lower heating power to high power, high \qpar{} scenarios for which the operational constraints will be more restrictive. The utility of the \Pin{} normalized trends is further motivated by robust correlations between the relevant plasma physics parameters, \Tet, \alphat, \cNe, and the D fueling and Ne seeding fluxes, \GamD{} and \GamNe, as shown in figure \ref{fig:phys_params_to_actuators}. Such correlations between \GamD{} and \Tet{} have already been demonstrated in JET experiments reconciling pump conductance differences between different divertor geometries \cite{lomanowski2022experimental}, while the fueling-to-impurity valve flux ratios have typically been employed in experiments to approximate the impurity concentration \cite{henderson2021parameter,kallenbach2024divertor} (e.g., $c_{\text{Ne}}\propto$ \GamNe/(\GamD+\GamNe)). The correlations between the physics parameter and valve fluxes are likely to depend on the fueling and seeding locations, whereas in figure \ref{fig:phys_params_to_actuators} we only evaluate the correlation for a single fueling and Ne seeding valve location.  The ability to map the key physics parameters to fueling and heating actuators potentially enables projections of \GamD{} and \GamNe{} requirements for a high power, detached, QCE scenario. We use the SOLPS datasets to illustrate such an approach as follows.

\begin{figure*}[t]
    \centering
    \includegraphics[width=0.95\linewidth]{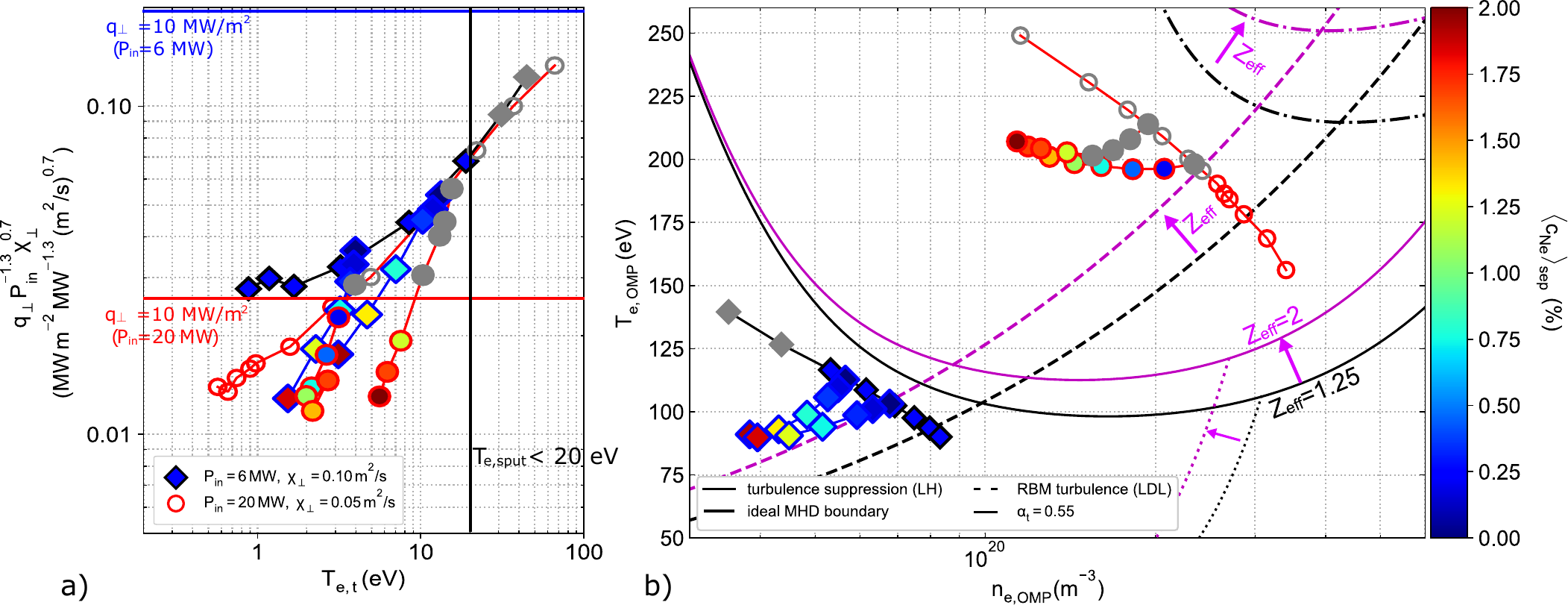}
    \caption{a) Dependence of normalized \qdep{} on \Tet{} with the nominal \Tet{}$=$ 20 eV peak flux tube sputtering threshold for the H-mode-like \Pin{}$=$20 MW, \chiperp{}$=$ 0.05 $\mathrm{m^2/s}$ cases and L-mode-like \Pin{}$=$ 6 MW, \chiperp{}$=$ 0.10 $\mathrm{m^2/s}$ cases. Also shown is the \qdep$=\mathrm{10 \: MW \: m^{-2}}$ threshold at the peak flux tube for the H-mode-like and L-mode-like scans. All L-mode-like cases are below \qdep$=\mathrm{10 \: MW \: m^{-2}}$.  b) The projected SPARC SepOS boundaries and \alphat{}$=$ 0.55 contours calculated assuming \Zeff{}$=$ 1.25 (black) and \Zeff{}$=$ 2 (magenta) for the full field $B_{\text{t}}=$12.2 T, $I_{\text{p}}=$8.7 MA scenario with the L-mode-like and H-mode-like density scans and Ne seeding branches corresponding to a). The marker fill color of the Ne branches indicates \cNe. L-mode-like and H-mode-like cases which are either above \qdep$=\mathrm{10 \: MW \: m^{-2}}$ or above \Tet{}$=$ 20 eV are grayed out.}
    \label{fig:SepOS_low_high_Pin}
\end{figure*}

In existing machines, L-mode operations at relatively low heating power are especially useful for informing detachment access conditions, characterizing and quantifying dissipation loss channels, as well as providing more benign ELM-free steady-state plasmas ideally suited for edge plasma diagnostics. As shown in figure \ref{fig:fmomfpwr}, the self-similar volumetric momentum and power loss factors are independent of power, and, although difficult to measure in H-mode, we expect these trends, and the other parameters of interest which exhibit strong \Tet{} dependence, to retain the self-similar characteristics independent of confinement regime. We therefore start by reproducing figure \ref{fig:Idiv_neomp_Tet_Ne}.c and \ref{fig:sepos_Ne} while limiting the comparison to the 12 T 8.7 MA \Pin{}$=$ 6 MW, \chiperp{}$=$ 0.1 $\mathrm{m^2/s}$ scenario which we consider to be L-mode-like, and the 12 T 8.7 MA H-mode-like scenario with \Pin{}$=$ 20 MW, \chiperp{}$=$ 0.05 $\mathrm{m^2/s}$, as shown in figure \ref{fig:SepOS_low_high_Pin}. As already noted, the factor of two reduction in \chiperp{} from L-mode to H-mode is approximately consistent with the observed factor of two reduction in \lamq{} in L-mode vs H-mode scenarios in AUG \cite{sieglin2017density}. 

To identify restricted operational points on the basis of exceeding \qdep{} material limits, we will for the present analysis adopt the typically used \qdep{}$<$ 10 $\mathrm{MW/m^2}$ constraint for W PFCs. We recognize that shot budgets for the SPARC W divertor will be evaluated by sophisticated workflows involving 3D modeling of thermal gradients and local stress concentrations in combination with thermal excursions above the W recrystallization temperature at the PFC surface. For simplicity we will consider simulation cases with \qdep{}$\gtrsim$ 10 $\mathrm{MW/m^2}$ at the peak flux tube to lie in a restricted part of operational space, with the assumption that such plasma solutions will significantly limit the discharge duration and/or will require additional mitigation techniques such as strike point sweeping.

Furthermore, since we are already using \Tet{} as the convenient divertor physics parameter connecting the PE considerations to the SepOS, we can impose a simple W sputtering constraint with a \Tet{}$\lesssim$ 20 eV requirement on the peak flux tube to prevent excessive physical sputtering. This is informed by operations with the JET W divertor and AUG W wall \cite{van2013tungsten}. We apply it here only as a rough measure, since the effective sputtering yields exhibit a strong dependence on the presence of intrinsic and extrinsic impurities and their concentrations \cite{eckstein2002calculated}. Additionally, in the presence of even small ELMs characteristic of the QCE regime, the W erosion could be intra-ELM dominated, therefore imposing an effectively lower ELM-averaged \Tet{} sputtering limit in H-mode scenarios. Given sufficiently high particle fluxes, W sputtering on flux tubes outboard of the peak heat flux tube with \Tet{}$\gtrsim$ 20 eV could also be significant even if \Tet{}$\lesssim$ 20 eV on the peak flux tube. While more detailed W net erosion projections for SPARC are currently under development, we proceed with the simple \Tet{}$\lesssim$ 20 eV sputtering threshold for illustration. 

Applying these constraints in figure \ref{fig:SepOS_low_high_Pin} yields a relatively less restricted operational space for the L-mode-like density and \cNe{} scans, as one might reasonably expect. The important point of the comparison is that the normalized \qdep\Pin$^{-1.3}$\chiperp$^{0.7}$ trends enable a projection of the accessible operational points at higher power from more readily established low power trends with higher safety margin. The commensurate \GamD{} and \GamNe{} requirements for achieving sufficiently low \Tet{} for detachment and suppression of W sputtering can also be extrapolated through the trends shown in figure \ref{fig:phys_params_to_actuators}. Thus, in this illustrative example, the remaining accessible operational points for the \Pin{}$=$ 20 MW H-mode-like scans (i.e., those points in figure \ref{fig:SepOS_low_high_Pin} that are not grayed out) are at i) high \cNe, with potentially more challenging access to QCE due to the significant \neomp{} reduction; ii) high \neomp{} and therefore high \alphat{} accessed through high fueling, but at low \cNe; and iii) scenarios in-between i) and ii) with partial recovery of the \neomp{} reduction with additional fueling, as shown in figure \ref{fig:alphat_Zeff}.c (but not included in figure \ref{fig:SepOS_low_high_Pin} for clarity). While both scenarios i) and ii) lead to sufficient \qdep{} reduction on the peak flux tube, the ELM regime for the high \cNe{} and lower \neomp{} operational points remains uncertain due to a lack of an experimental basis for the SepOS in high radiative fraction scenarios. Input from existing experiments on JET and AUG is limited to the extent that the highly radiative, partially detached regime with impurity radiation concentrated in the SOL can be accurately distinguished from the XPR regime. We consider the XPR regime as a soft operational limit but which can nevertheless have a significant impact on the ELM behavior that is distinct from ELM behavior in the partially detached scenarios, as observed from JET and AUG experiments \cite{bernert2017power,gloggler2019characterisation,bernert2021x,bernert2023x}. This also further motivates comparisons between SPARC XPR access conditions and the AUG XPR interpretive studies \cite{pan2023solps} in order to gauge the operational window for partially detached high $f_{\text{rad}}$ scenarios on SPARC. Also worth noting is that a lower \Tet{} sputtering threshold of 10 eV instead of 20 eV would not significantly restrict the operational space for the \Pin{}$=$ 20 MW H-mode-like scans, but would further restrict the L-mode-like operational points that are below \qdep{}$=$ 10 $\mathrm{MW/m^2}$ but with \Tet{} $>$ 10 eV. Lastly, despite the \qdep{} reduction below 10 $\mathrm{MW/m^2}$ on the peak flux tube for the higher density unseeded cases (figure \ref{fig:SepOS_low_high_Pin}.a), the maximum \qdep{} outboard of the peak flux tube for these cases remains elevated at 10-12 $\mathrm{MW/m^2}$, indicating that the neutral-plasma dissipation is most effective in the first few flux tubes in the outer horizontal target configuration. In the Ne seeded cases the radiative dissipation helps to further reduce \qdep{} outboard of the peak flux tube. The addition of drifts may also impact the target \qdep{} footprint, hence further work is needed to characterize the evolution of the \qdep{} profile on both the outer horizontal and outer vertical target configurations, similar to interpretive studies on JET \cite{groth2015divertor,moulton2018neutral}. 

To avoid over interpretation of the above SOLPS results, the accuracy of the presented predictions  isn't as important as the basic principle of the approach, i.e., establishing lower power trends using appropriate observables in experiment and extrapolating to higher power, more restricted operational points through correlations between the key physics parameters and fueling and heating actuators.

\section{\label{sec:Discussion}Discussion}

Analysis of the SOLPS density and Ne scans in the context of the PE-SepOS framework suggests a balance or compromise is needed for accessing the QCE regime in SPARC while ensuring sufficient heat flux mitigation and W erosion suppression via extrinsic Ne seeding. Achieving sufficiently high \neomp{} via fueling in order to exceed the \alphat{} $>$ 0.55 empirical threshold is made more challenging in the presence of significant \cNe{} due to the power limitation effect which reduces target recycling and plasma pressure. Through parallel pressure balance, this results in significant \neomp{} reduction thus limiting a further rise in, or even decreasing, \alphat, depending on the interplay of \Zeff{} and \Teomp{} and level of recycling. However, recent demonstration of a high performance semi-detached Ne seeded scenario during the JET DTE3 experimental campaign shows that at sufficiently high heating power, the ELM size and frequency can be reduced with increasing Ne concentration including an ELM-free regime at $c_{\text{Ne}}=$1.7 \% (estimated at the pedestal top) while maintaining good confinement in both D-D and D-T plasmas \cite{Giroud2025JETDTE3}. Similar to earlier lower heating power Ne seeding experiments on JET with the Be-W wall \cite{gloggler2019characterisation}, these high triangularity ITER-like shaped scenarios exhibit a significant \neomp{} reduction consistent with the SOLPS modeling results presented here. This is accompanied by a pronounced reduction in the pedestal density (50\% at $c_{\text{Ne}}=$1.7\%), although balanced by a strong increase in the pedestal $T_{e}$ and $T_{i}$, yielding an increasing pedestal pressure with $c_{\text{Ne}}$, a result which was achieved at the highest available heating power.  

Detailed analysis of the pedestal stability of these discharges in \cite{nystrom2025effect} indicates similarities to the QCE regime in terms of destabilizing infinite-n ballooning modes at the pedestal foot. Moreover, an initial evaluation of the separatrix parameters suggests an overall rise of \alphat{} with $c_{\text{Ne}}$ exceeding the \alphat{}$=$ 0.55 QCE threshold at $c_{\text{Ne}}=$1.7 \%, albeit with significant measurement uncertainty. Similar qualitative behavior is observed in the SOLPS SPARC datasets in figure \ref{fig:alphat_Zeff}.a for the highest density Ne seeding branch which first leads to a decrease in \alphat{} and subsequent recovery at higher \Zeff. Resolving the impact of drifts is an important next step in the modeling as it could lead to significant Ne redistribution, consequently impacting the presented Ne seeding trends.

The regulation of \neomp{} with $c_{\text{Ne}}$ observed in the JET Ne seeding experiments is also considered a critical aspect in accessing lower collisionality pedestal top conditions leading to a rise in pedestal top pressure \cite{nystrom2025effect}. Therefore, high \cNe{} scenarios which exhibit favorable small ELM and ELM-free behavior can potentially enable higher pedestal performance, as well as facilitate access to a lower density detached scenario with more favorable H-mode access (see \cite{hughes2025high}) with an operational point closer to the targeted SPARC Primary Reference Discharge (PRD) \cite{creely2023sparc} Greenwald fraction ($f_{\text{GW,ped}}\approx 0.34$). 

\section{\label{sec:Conclusions}Conclusions}

In this contribution we developed a framework for linking power exhaust constraints to the projected separatrix operational space for SPARC (i.e., the PE-SepOS). The tradeoffs that emerge in evaluating small-ELM access conditions and detachment criteria with Ne seeding demonstrate the necessity and utility of the PE-SepOS workflow for finding integrated power exhaust solutions. 

We mapped the upstream and divertor target plasma parameters from the SOLPS density scan at the peak SOL flux tube to \Tet, which we consider the primary physics parameter for characterizing detachment. Using heating-power and transport-coefficient normalization factors derived from 2PM-like \qpar{} scaling, we obtained self-similar trends that unify the particle detachment rollover and key parameters like \neomp, \Teomp, \qdep{} and the net divertor dissipation loss factors \fmom{} and \fpwr. Recasting the density scans onto the SepOS defined in terms of \neomp, \Teomp{} and \alphat, we find that the particle flux rollover detachment reference point consistently intersects the QCE access \alphat{} $\approx$ 0.55 condition for the entire range of considered heating powers and transport assumptions with fixed \Dperp/\chiperp. This relationship emerges as a consequence of a strong correlation between \Tet{} and \alphat$\propto n_{e\mathrm{,OMP}}/T_{e\mathrm{,OMP}}^2$ for fixed \qcyl{} and \Zeff. Using a flux-tube averaged Spitzer-H{\"a}rm power balance approach to recover the decreasing \Teomp{} trend with increasing \neomp, we find that the flux-limited parallel electron heat conductivity model in SOLPS leads to higher \Teomp{} at lower densities compared to the more simplified Spitzer-H{\"a}rm model used to interpret \Teomp{} in experiments. With this quantitative understanding of both \neomp{} and \Teomp{} trends, we derived \alphat$\propto  T_{e\mathrm{,t}}^{-1/2}(1-f_{\mathrm{pwr}})/(1-f_{\mathrm{mom}})$, thus articulating the causal relationship between \alphat{} and the net momentum and power losses driven by neutral-plasma interactions near the outer divertor target. Hence, for fixed plasma shaping in steady-state unseeded scenarios, we demonstrate a strong link between the key SPARC SepOS parameters and the normalized divertor target quantities, such that a given \alphat{} value will have a corresponding divertor state that cannot be decoupled via gas fueling and heating actuators.

In extending the SOLPS unseeded density scans with Ne concentration scans at fixed gas fueling, we observed a significant reduction of \neomp{} by up to 50\% at \cNe$=$2\%. Such a pronounced \neomp{} reduction is consistent with Ne seeded scenarios on JET as well as SOLPS modelling supporting the ITER divertor physics basis. This arises due to the \enquote*{power limitation} effect, which leads to a reduction of neutral recycling at the divertor targets due to an increase in seeded impurity radiation, energy which would otherwise be available to sustain the neutral excitation, ionization and dissociation atomic recycling processes. The reduced recycling and hence the target pressure is then propagated upstream through momentum balance leading to a decrease in \neomp. 

In contrast to gas fueled detachment obtained via neutral-plasma momentum and power dissipation with increasing \neomp{} and \alphat, mitigation of excessive heat fluxes via Ne radiation results in a largely constant or decreasing \alphat{} trend with increasing \cNe, potentially impacting access to QCE regimes in highly radiative partially detached scenarios. However, since the SepOS framework has so far not been demonstrated in impurity seeded highly radiative regimes, extrapolation of the empirical \alphat{} $>$ 0.55 QCE access condition to the high \cNe{} scenarios remains uncertain. Recent demonstrations of an ELM-free high performance partially detached integrated scenario with significant Ne seeding in JET supports high $c_{\text{Ne}}$ as a possible alternate route to QCE-like behavior via increasing \Zeff{} which potentially balances the reduction in \neomp. The nuanced interplay of \neomp, \Teomp{} and \Zeff{} on \alphat{} emerging from the SOLPS analysis needs further evaluation in drift-activated simulations to quantify the impact on Ne redistribution in SPARC, which could provide further insight into the \alphat{} trends.  Nevertheless, the tradeoffs in gas fueling and Ne seeding illustrated through the dedicated SOLPS scans for SPARC highlight the important role of \neomp{} as an optimization parameter in the PE-SepOS framework, and more broadly in the context of global scenario optimization. This work motivates renewed further efforts in characterizing ELM behavior in the presence of significant extrinsic impurity content, and in particular by introducing impurity seeding in established gas fueled QCE operational points on existing devices while avoiding full transition to the X-point radiating regime, which we consider to be a soft limit to the partially-detached operational space explored in this work. 

A key insight that emerges from the PE-SepOS predictions for SPARC is the robustness of the self-similar trends with respect to \Tet{}. These trends extend beyond the unseeded density scans to the Ne seeding \enquote*{branches}, exhibiting similar unified trends independent of heating power and cross-field transport assumptions. The practical significance of this finding is the potential for extrapolation from low power trends established in scenarios with more relaxed safety margins to more restricted high power operational points. This scaling up capability is further motivated by robust correlations between the actuators \GamD, \GamNe{} and \Psep{} and the key power exhaust physics parameters \Tet{} and \cNe, although different fueling/seeding locations (and thus differences in fueling efficiency and/or impurity screening) could introduce further variability not captured in this work. Assuming nominal \qdep$=\mathrm{10 \: MW \: m^{-2}}$ limits on the peak flux tube --- a simplified illustrative limit for W PFCs relative to the high fidelity recrystallization kinetics and thermal gradient modeling workflows being developed to guide SPARC pulse planning --- the PE-SepOS projections indicate accessible operational points with \qdep$<\mathrm{10 \: MW \: m^{-2}}$ and \Tet$<$10 eV for \Psep$<$20-29 MW with \alphat$>$0.55 achieved through high density, moderate \cNe{} scenarios. A more restrictive operational space is found at lower densities assuming \lamq{} in the range 0.3-0.6 mm. Alternatively, similar reductions in \qdep{} are achieved via high \cNe{} scenarios at reduced densities, with the remaining uncertainty being the ELM characteristics in this regime and the \alphat{} balance between reduction of \neomp{} via power limitation vs the increase in \Zeff{} at higher \cNe. 

Given the overall importance of \Tet{} and \alphat{} in facilitating the PE-SepOS framework, developing measurement capability aiming at quantitative recovery of these parameters will be key to make use of the observed self-similar behavior for informing SPARC operations. Lastly, these projections are currently independent of pedestal and global performance considerations. In ongoing work, further consideration of pedestal-SOL coupling leveraging the recently developed EPED-SOLPS workflows \cite{wilcox2026towards} aims at imposing pedestal performance and impurity enrichment criteria on the PE-SepOS framework (i.e., towards the power exhaust and pedestal constrained separatrix operational space, or the PEP-SepOS integrated framework).   

\textbf{ }

\section*{Acknowledgments}
This manuscript has been authored by UT-Battelle, LLC, with the US Department of Energy. The work was supported by the U.S. Department of Energy’s Office of Fusion Energy Sciences Compact Toroidal Concepts program, namely \enquote{Studies of Fusion Pilot Plant Physics in SPARC Early Campaigns and Q greater than 1 Plasmas}, by the Innovation Network for Fusion Energy (INFUSE), and by Commonwealth Fusion Systems. This research used resources of the Oak Ridge National Laboratory Research Cloud. The results are obtained with the help of the EIRENE package (see www.eirene.de) including the related code, data and tools \cite{ryter2014experimental}. Oak Ridge National Laboratory is operated for the DOE by UT-Battelle, LLC under contract DE-AC05-00OR22725. The authors dedicate this work to the memory of Professor Peter Stangeby, whose insight was invaluable to this study.

\balance 

\bibliographystyle{iopart-num}
\bibliography{references}

@PREAMBLE{
 "\providecommand{\noopsort}[1]{}" 
 # "\providecommand{\singleletter}[1]{#1}%" 
}

@article{lore2024evaluation,
  title={Evaluation of SPARC divertor conditions in H-mode operation using SOLPS-ITER},
  author={Lore, Jeremy D and Park, Jae-Sun and Eich, Thomas and Kuang, Adam Q and Reinke, Matthew L and De Pascuale, Sebastian and Lomanowski, Bart and Creely, Alex and Canik, John M},
  journal={Nuclear Fusion},
  volume={64},
  number={12},
  pages={126054},
  year={2024},
  publisher={IOP Publishing}
}

@article{sieglin2017density,
  title={Density dependence of SOL power width in ASDEX upgrade L-Mode},
  author={Sieglin, B and Eich, T and Faitsch, M and Herrmann, A and Nille, D and Scarabosio, A and ASDEX Upgrade Team and others},
  journal={Nuclear Materials and Energy},
  volume={12},
  pages={216--220},
  year={2017},
  publisher={Elsevier}
}

@article{eich2020turbulence,
  title={Turbulence driven widening of the near-SOL power width in ASDEX Upgrade H-Mode discharges},
  author={Eich, T and Manz, P and Goldston, RJ and Hennequin, Pascale and David, P and Faitsch, M and Kurzan, B and Sieglin, B and Wolfrum, E and ASDEX Upgrade Team and others},
  journal={Nuclear Fusion},
  volume={60},
  number={5},
  pages={056016},
  year={2020},
  publisher={IOP Publishing}
}

@book{stangeby2000plasma,
  title={The plasma boundary of magnetic fusion devices},
  author={Stangeby, Peter C},
  year={2000},
  publisher={CRC Press}
}

@article{stangeby2018basic,
  title={Basic physical processes and reduced models for plasma detachment},
  author={Stangeby, PC},
  journal={Plasma Physics and Controlled Fusion},
  volume={60},
  number={4},
  pages={044022},
  year={2018},
  publisher={IOP Publishing}
}

@article{lomanowski2019spectroscopic,
  title={Spectroscopic investigation of N and Ne seeded induced detachment in JET ITER-like wall L-modes combining experiment and EDGE2D modeling},
  author={Lomanowski, B and Carr, Matthew and Field, Anthony and Groth, Mathias and Jaervinen, AE and Lowry, Christopher and Meigs, AG and Menmuir, Sheena and O'Mullane, M and Reinke, ML and others},
  journal={Nuclear Materials and Energy},
  volume={20},
  pages={100676},
  year={2019},
  publisher={Elsevier}
}

@article{lomanowski2020interpretation,
  title={Interpretation of Lyman opacity measurements in JET with the ITER-like wall using a particle balance approach},
  author={Lomanowski, B and Groth, M and Coffey, Ivor and Karhunen, Juuso and Maggi, CF and Meigs, AG and Menmuir, Sheena and O’Mullane, M and others},
  journal={Plasma Physics and Controlled Fusion},
  volume={62},
  number={6},
  pages={065006},
  year={2020},
  publisher={IOP Publishing}
}

@article{lomanowski2023variation,
  title={Variation in the volumetric power and momentum losses in the JET-ILW scrape-off layer},
  author={Lomanowski, B and Park, JS and Aho-Mantila, L and Brix, M and Groth, M and Guillemaut, C and Lowry, C and Marsen, S and Meigs, A and Wischmeier, M and others},
  journal={Nuclear Materials and Energy},
  volume={35},
  pages={101425},
  year={2023},
  publisher={Elsevier}
}

@article{lomanowski2022experimental,
  title={Experimental study on the role of the target electron temperature as a key parameter linking recycling to plasma performance in JET-ILW},
  author={Lomanowski, B and Dunne, M and Vianello, N and Aleiferis, S and Brix, M and Canik, J and Carvalho, IS and Frassinetti, Lorenzo and Frigione, D and Garzotti, L and others},
  journal={Nuclear Fusion},
  volume={62},
  number={6},
  pages={066030},
  year={2022},
  publisher={IOP Publishing}
}

@article{lomanowski2023parameter,
  title={Parameter dependencies of the separatrix density in low triangularity L-mode and H-mode JET-ILW plasmas},
  author={Lomanowski, Bartosz and Rubino, Giulio and Uccello, Andrea and Dunne, M and Vianello, Nicola and Aleiferis, Spyridon and Canik, John and Carvalho, I and Corrigan, G and Frassinetti, Lorenzo and others},
  journal={Nuclear Fusion},
  volume={63},
  number={3},
  pages={036019},
  year={2023},
  publisher={IOP Publishing}
}

@article{park2024impact,
  title={Impact of gas injection location and divertor surface material on ITER fusion power operation phase divertor performance assessed with SOLPS-ITER},
  author={Park, Jae-Sun and Bonnin, Xavier and Pitts, Richard and Lore, Jeremy},
  journal={Nuclear Fusion},
  volume={64},
  number={3},
  pages={036002},
  year={2024},
  publisher={IOP Publishing}
}

@article{eich2021separatrix,
  title={The separatrix operational space of ASDEX Upgrade due to interchange-drift-Alfv{\'e}n turbulence},
  author={Eich, Thomas and Manz, Peter and ASDEX Upgrade Team and others},
  journal={Nuclear Fusion},
  volume={61},
  number={8},
  pages={086017},
  year={2021},
  publisher={IOP Publishing}
}

@article{eich2025separatrix,
  title={The separatrix operational space of next-step fusion experiments: from ASDEX Upgrade data to SPARC scenarios},
  author={Eich, Thomas and Body, Thomas and Faitsch, Michael and Grover, Ondrej and Miller, Marco Andres and Manz, Peter and Looby, Tom and Kuang, Adam Qingyang and Redl, Andreas and Reinke, Matt and others},
  journal={Nuclear Materials and Energy},
  volume={42},
  pages={101896},
  year={2025},
  publisher={Elsevier}
}

@article{hughes2025high,
  title={High confinement regimes on SPARC: operational conditions for access and avoidance},
  author={Hughes, JW and Rodriguez-Fernandez, P and Hubbard, AE and Battaglia, DJ and Miller, MA and Cavallaro, A and Howard, NT and Wilks, TM and Creely, AJ},
  journal={Nuclear Fusion},
  volume={65},
  number={5},
  pages={052001},
  year={2025},
  publisher={IOP Publishing}
}

@article{ryter2014experimental,
  title={Experimental evidence for the key role of the ion heat channel in the physics of the L--H transition},
  author={Ryter, F and Orte, L Barrera and Kurzan, B and McDermott, RM and Tardini, G and Viezzer, E and Bernert, M and Fischer, R and ASDEX Upgrade Team and others},
  journal={Nuclear Fusion},
  volume={54},
  number={8},
  pages={083003},
  year={2014},
  publisher={IOP Publishing}
}

@misc{eckstein2002calculated,
  title={Calculated sputtering, reflection and range values (IPP 9/132)},
  author={Eckstein, W},
  year={2002},
  publisher={Munich: Garching: Max-Planck-Institute f{\"u}r Plasmaphysik}
}

@article{faitsch2023analysis,
  title={Analysis and expansion of the quasi-continuous exhaust (QCE) regime in ASDEX Upgrade},
  author={Faitsch, Michael and Eich, T and Harrer, GF and Wolfrum, Elisabeth and Brida, Dominik and David, Pierre and Dunne, Mike and Gil, Lu{\'\i}s and Labit, Benoit and Stroth, Ulrich and others},
  journal={Nuclear Fusion},
  volume={63},
  number={7},
  pages={076013},
  year={2023},
  publisher={IOP Publishing}
}

@article{moulton2024super,
  title={Super-X and conventional divertor configurations in MAST-U ohmic L-mode; a comparison facilitated by interpretative modelling},
  author={Moulton, David and Harrison, JR and Xiang, L and Ryan, PJ and Kirk, A and Verhaegh, Kevin and Wijkamp, Tijs A and Federici, Fabio and Clark, JG and Lipschultz, B},
  journal={Nuclear Fusion},
  volume={64},
  number={7},
  pages={076049},
  year={2024},
  publisher={IOP Publishing}
}

@article{day1996effect,
  title={The effect of heat flux limiting on divertor fluid models},
  author={Day, M and Merriman, B and Najmabadi, F and Conn, RW},
  journal={Contributions to Plasma Physics},
  volume={36},
  number={2-3},
  pages={419--423},
  year={1996},
  publisher={Wiley Online Library}
}

@article{fundamenski2005parallel,
  title={Parallel heat flux limits in the tokamak scrape-off layer},
  author={Fundamenski, W},
  journal={Plasma physics and controlled fusion},
  volume={47},
  number={11},
  pages={R163},
  year={2005},
  publisher={IOP Publishing}
}

@article{silvagni2025separatrix,
  title={The separatrix electron density in JET, ASDEX upgrade and alcator C-Mod H-mode plasmas: A common evaluation procedure and correlation with engineering parameters},
  author={Silvagni, D and Grover, O and Stagni, A and Hughes, JW and Miller, MA and Lomanowski, B and Ciraolo, G and Dunne, M and Eich, T and Frassinetti, L and others},
  journal={Nuclear Materials and Energy},
  volume={42},
  pages={101867},
  year={2025},
  publisher={Elsevier}
}

@article{goldston2017new,
  title={A new scaling for divertor detachment},
  author={Goldston, Robert James and Reinke, ML and Schwartz, JA},
  journal={Plasma Physics and Controlled Fusion},
  volume={59},
  number={5},
  pages={055015},
  year={2017},
  publisher={IOP Publishing}
}

@article{pitts2019physics,
  title={Physics basis for the first ITER tungsten divertor},
  author={Pitts, Richard A and Bonnin, Xavier and Escourbiac, Fr{\'e}d{\'e}ric and Frerichs, Heinke and Gunn, JP and Hirai, Takeshi and Kukushkin, AS and Kaveeva, Elizaveta and Miller, MA and Moulton, David and others},
  journal={Nuclear Materials and Energy},
  volume={20},
  pages={100696},
  year={2019},
  publisher={Elsevier}
}

@article{bernert2017power,
  title={Power exhaust by SOL and pedestal radiation at ASDEX Upgrade and JET},
  author={Bernert, M and Wischmeier, M and Huber, A and Reimold, F and Lipschultz, B and Lowry, C and Brezinsek, S and Dux, R and Eich, T and Kallenbach, A and others},
  journal={Nuclear Materials and Energy},
  volume={12},
  pages={111--118},
  year={2017},
  publisher={Elsevier}
}

@article{bernert2023x,
  title={The X-Point radiating regime at ASDEX Upgrade and TCV},
  author={Bernert, M and Wiesen, S and F{\'e}vrier, O and Kallenbach, A and Koenders, JTW and Sieglin, B and Stroth, U and Bosman, TOSJ and Brida, D and Cavedon, M and others},
  journal={Nuclear Materials and Energy},
  volume={34},
  pages={101376},
  year={2023},
  publisher={Elsevier}
}

@article{bernert2021x,
  title={X-point radiation, its control and an ELM suppressed radiating regime at the ASDEX Upgrade tokamak},
  author={Bernert, M and Janky, F and Sieglin, B and Kallenbach, A and Lipschultz, B and Reimold, F and Wischmeier, M and Cavedon, M and David, P and Dunne, MG and others},
  journal={Nuclear Fusion},
  volume={61},
  number={2},
  pages={024001},
  year={2021},
  publisher={IOP Publishing}
}

@article{kallenbach2024divertor,
  title={Divertor enrichment of recycling impurity species (He, N2, Ne, Ar, Kr) in ASDEX Upgrade H-modes},
  author={Kallenbach, A and Dux, R and Henderson, SS and Tantos, C and Bernert, M and Day, C and McDermott, RM and Rohde, V and Zito, A and ASDEX Upgrade Team},
  journal={Nuclear Fusion},
  volume={64},
  number={5},
  pages={056003},
  year={2024},
  publisher={IOP Publishing}
}

@article{henderson2021parameter,
  title={Parameter dependencies of the experimental nitrogen concentration required for detachment on ASDEX Upgrade and JET},
  author={Henderson, Stuart S and Bernert, M and Giroud, C and Brida, D and Cavedon, M and David, P and Dux, R and Harrison, JR and Huber, A and Kallenbach, A and others},
  journal={Nuclear Materials and Energy},
  volume={28},
  pages={101000},
  year={2021},
  publisher={Elsevier}
}

@article{van2013tungsten,
  title={Tungsten divertor erosion in all metal devices: Lessons from the ITER like wall of JET},
  author={Van Rooij, GJ and Coenen, JW and Aho-Mantila, Leena and Brezinsek, S and Clever, M and Dux, R and Groth, M and Krieger, K and Marsen, S and Matthews, GF and others},
  journal={Journal of Nuclear Materials},
  volume={438},
  pages={S42--S47},
  year={2013},
  publisher={Elsevier}
}

@article{pan2023solps,
  title={SOLPS-ITER simulations of an X-point radiator in the ASDEX Upgrade tokamak},
  author={Pan, O and Bernert, M and Lunt, T and Cavedon, M and Kurzan, B and Wiesen, S and Wischmeier, M and Stroth, U and Upgrade Team, the ASDEX},
  journal={Nuclear Fusion},
  volume={63},
  number={1},
  pages={016001},
  year={2023},
  publisher={IOP Publishing}
}

@article{gloggler2019characterisation,
  title={Characterisation of highly radiating neon seeded plasmas in JET-ILW},
  author={Gl{\"o}ggler, S and Wischmeier, M and Fable, E and Solano, ER and Sertoli, M and Bernert, M and Calabr{\`o}, G and Chernyshova, M and Huber, A and Kowalska-Strz{\k{e}}ciwilk, E and others},
  journal={Nuclear Fusion},
  volume={59},
  number={12},
  pages={126031},
  year={2019},
  publisher={IOP Publishing}
}

@article{eich2013scaling,
  title={Scaling of the tokamak near the scrape-off layer H-mode power width and implications for ITER},
  author={Eich, Thomas and Leonard, AW and Pitts, RA and Fundamenski, W and Goldston, Robert James and Gray, TK and Herrmann, A and Kirk, A and Kallenbach, A and Kardaun, O and others},
  journal={Nuclear fusion},
  volume={53},
  number={9},
  pages={093031},
  year={2013},
  publisher={IOP Publishing and International Atomic Energy Agency}
}

@article{eich2011PRL,
  title = {Inter-ELM Power Decay Length for JET and ASDEX Upgrade: Measurement and Comparison with Heuristic Drift-Based Model},
  author = {Eich, T. and Sieglin, B. and Scarabosio, A. and Fundamenski, W. and Goldston, R. J. and Herrmann, A.},
  collaboration = {ASDEX Upgrade Team},
  journal = {Phys. Rev. Lett.},
  volume = {107},
  issue = {21},
  pages = {215001},
  numpages = {4},
  year = {2011},
  month = {Nov},
  publisher = {American Physical Society},
  doi = {10.1103/PhysRevLett.107.215001},
  url = {https://link.aps.org/doi/10.1103/PhysRevLett.107.215001}
}

@article{dunne2024quasi,
  title={Quasi-continuous exhaust operational space},
  author={Dunne, M and Faitsch, Michael and Radovanovic, Lidija and Wolfrum, Elisabeth and ASDEX Upgrade Team},
  journal={Nuclear Fusion},
  volume={64},
  number={12},
  pages={124003},
  year={2024},
  publisher={IOP Publishing}
}

@article{park2024actuator,
  title={Full time-dependent SOLPS-ITER simulation of the SPARC tokamak: actuator design for particle and divertor condition control},
  author={Park, Jae-Sun and Lore, Jeremy D and Reinke, Matthew and Kuang, Adam Q and De Pascuale, Sebastian and Creely, Alex},
  journal={Nuclear Fusion},
  volume={64},
  number={7},
  pages={076036},
  year={2024},
  publisher={IOP Publishing}
}

@article{wilcox2026towards,
  title={Towards self-consistent integrated modeling of the tokamak pedestal, scrape-off layer, and divertor using SOLPS-ITER and EPED},
  author={Wilcox, Robert S and Canik, John M and Park, JM and Snyder, Philip B and Shafer, Morgan W and De Pascuale, Sebastian and Easley, Davis and Leonard, Anthony W and Lore, Jeremy D and McLean, Adam G and others},
  journal={Nuclear Fusion},
  year={2026}
}

@article{park2026DsChis,
  title={Transport driven variation of SOL decay lengths and divertor power sharing in SPARC SOLPS simulations},
  author={Park, Jae-Sun and Lore, Jeremy D and Lomanowski, Bart and Eich, Thomas and Body, Tom},
  journal={in preparation},
  year={2026}
}

@misc{tom_body_2024_13820268,
  author       = {Tom Body and
                  Christoph Hasse and
                  Audrey Saltzman and
                  Allen Wang and
                  IsaacSavona and
                  Oak Nelson and
                  Tom Looby},
  title        = {cfs-energy/cfspopcon: v7.0.2},
  month        = sep,
  year         = 2024,
  publisher    = {Zenodo},
  version      = {v7.0.2},
  doi          = {10.5281/zenodo.13820268},
  url          = {https://doi.org/10.5281/zenodo.13820268},
}

@article{delabie2026empirical,
  title={Empirical scaling of the LH threshold power for metal wall tokamaks using a multi-device database},
  author={Delabie, Ephrem and Solano, Emilia R and Hughes, Jerry W and Maggi, Costanza F and Ryter, Francois and Birkenmeier, Gregor and Carvalho, Pedro and Cavedon, Marco and Chernyshova, Maryna and David, Pierre and others},
  journal={Nuclear Fusion},
  year={2026}
}

@article{schmidtmayr2018investigation,
  title={Investigation of the critical edge ion heat flux for LH transitions in Alcator C-Mod and its dependence on BT},
  author={Schmidtmayr, Matthias and Hughes, JW and Ryter, F and Wolfrum, Elisabeth and Cao, N and Creely, AJ and Howard, N and Hubbard, AE and Lin, Y and Reinke, ML and others},
  journal={Nuclear Fusion},
  volume={58},
  number={5},
  pages={056003},
  year={2018},
  publisher={IOP Publishing}
}

@article{stangeby2015identifying,
  title={Identifying the location of the OMP separatrix in DIII-D using power accounting},
  author={Stangeby, Peter C and Canik, John M and Elder, JD and Lasnier, CJ and Leonard, AW and Eldon, D and Makowski, MA and Osborne, TH and Grierson, Brian A},
  journal={Nuclear Fusion},
  volume={55},
  number={9},
  pages={093014},
  year={2015},
  publisher={IOP Publishing}
}

@article{sun2020empirical,
  title={Empirical study of gradient lengths ratio $\eta$ e in the near SOL region in ASDEX Upgrade tokamak},
  author={Sun, HJ and Wolfrum, E and Eich, T and Kallenbach, A and Schneider, P and Kurzan, B and Stroth, U and ASDEX Upgrade Team},
  journal={Plasma Physics and Controlled Fusion},
  volume={62},
  number={2},
  pages={025005},
  year={2020},
  publisher={IOP Publishing}
}

@manual{coster_solps_manual,
  author       = {D. P. Coster},
  title        = {{SOLPS Manual}},
  organization = {Max Planck Institute for Plasma Physics},
  year         = {n.d.},
  url          = {http://solps-mdsplus.aug.ipp.mpg.de:8080/solps/Documentation/solps.pdf},
  note         = {Accessed: 2026-05-05}
}

@inproceedings{Giroud2025JETDTE3,
  author       = {Carine Giroud},
  title        = {High performance ELM-free semi-detached scenario sustained at high-current in JET DTE3},
  booktitle    = {Proceedings of the 30th IAEA Fusion Energy Conference (FEC 2025)},
  year         = {2025},
  address      = {Chengdu, China},
  organization = {International Atomic Energy Agency (IAEA)},
  note         = {Conference synopsis (presentation), UKAEA},
}

@article{nystrom2025effect,
  title={Effect of neon seeding on the pedestal structure and stability in the JET-ILW deuterium JET-ITER baseline scenario using both ideal and resistive MHD},
  author={Nystr{\"o}m, Hampus and Giroud, Carine and Frassinetti, Lorenzo and Menmuir, Sheena and Litherland-Smith, Evie and Fontdecaba Climent, Josep Maria and Kos, Domagoj and Huang, Zhouji and Boboc, Alexandru and Macdonald, James and others},
  journal={Nuclear Fusion},
  year={2025}
}

@article{reimold2015divertor,
  title={Divertor studies in nitrogen induced completely detached H-modes in full tungsten ASDEX Upgrade},
  author={Reimold, Felix and Wischmeier, M and Bernert, M and Potzel, S and Kallenbach, A and M{\"u}ller, Hans Werner and Sieglin, B and Stroth, U and ASDEX Upgrade Team},
  journal={Nuclear Fusion},
  volume={55},
  number={3},
  pages={033004},
  year={2015},
  publisher={IOP Publishing}
}

@article{matthews2011jet,
  title={JET ITER-like wall—overview and experimental programme},
  author={Matthews, GF and Beurskens, M and Brezinsek, S and Groth, M and Joffrin, E and Loving, A and Kear, M and Mayoral, ML and Neu, R and Prior, P and others},
  journal={Physica Scripta},
  volume={2011},
  number={T145},
  pages={014001},
  year={2011}
}

@article{brezinsek2015plasma,
  title={Plasma-surface interaction in the Be/W environment: Conclusions drawn from the JET-ILW for ITER},
  author={Brezinsek, Sebastijan and JET-EFDA contributors and others},
  journal={Journal of nuclear materials},
  volume={463},
  pages={11--21},
  year={2015},
  publisher={Elsevier}
}

@article{loarte2011high,
  title={High confinement/high radiated power H-mode experiments in Alcator C-Mod and consequences for International Thermonuclear Experimental Reactor (ITER) QDT= 10 operation},
  author={Loarte, A and Hughes, JW and Reinke, ML and Terry, JL and LaBombard, B and Brunner, D and Greenwald, M and Lipschultz, B and Ma, Y and Wukitch, S and others},
  journal={Physics of Plasmas},
  volume={18},
  number={5},
  year={2011},
  publisher={AIP Publishing}
}

@article{loarte2007chapter,
  title={Chapter 4: Power and particle control},
  author={Loarte, Alberto and Lipschultz, Bruce and Kukushkin, AS and Matthews, GF and Stangeby, PC and Asakura, Nobuyuki and Counsell, GF and Federici, Gianfranco and Kallenbach, Arne and Krieger, K and others},
  journal={Nuclear Fusion},
  volume={47},
  number={6},
  pages={S203--S263},
  year={2007}
}

@article{kallenbach2013impurity,
  title={Impurity seeding for tokamak power exhaust: from present devices via ITER to DEMO},
  author={Kallenbach, Arne and Bernert, Matthias and Dux, R and Casali, Livia and Eich, Thomas and Giannone, Louis and Herrmann, Albrecht and McDermott, R and Mlynek, Alexander and M{\"u}ller, Hans Werner and others},
  journal={Plasma Physics and Controlled Fusion},
  volume={55},
  number={12},
  pages={124041},
  year={2013},
  publisher={IOP Publishing}
}

@article{reinke2017heat,
  title={Heat flux mitigation by impurity seeding in high-field tokamaks},
  author={Reinke, ML},
  journal={Nuclear Fusion},
  volume={57},
  number={3},
  pages={034004},
  year={2017},
  publisher={IOP Publishing}
}

@article{kallenbach2019neutral,
  title={Neutral pressure and separatrix density related models for seed impurity divertor radiation in ASDEX Upgrade},
  author={Kallenbach, A and Bernert, M and Dux, R and Eich, T and Henderson, SS and P{\"u}tterich, T and Reimold, F and Rohde, V and Sun, HJ and ASDEX Upgrade Team and others},
  journal={Nuclear Materials and Energy},
  volume={18},
  pages={166--174},
  year={2019},
  publisher={Elsevier}
}

@article{kukushkin2011finalizing,
  title={Finalizing the ITER divertor design: The key role of SOLPS modeling},
  author={Kukushkin, AS and Pacher, HD and Kotov, Vladislav and Pacher, GW and Reiter, Detlev},
  journal={Fusion engineering and design},
  volume={86},
  number={12},
  pages={2865--2873},
  year={2011},
  publisher={Elsevier}
}

@article{lore2022high,
  title={High gas throughput SOLPS-ITER simulations extending the ITER database to strong detachment},
  author={Lore, Jeremy D and Bonnin, X and Park, J-S and Pitts, RA and Stangeby, PC},
  journal={Nuclear Fusion},
  volume={62},
  number={10},
  pages={106017},
  year={2022},
  publisher={IOP Publishing}
}

@article{cowley2023novel,
  title={Novel SOLPS-ITER simulations of X-point target and snowflake divertors},
  author={Cowley, C and Kuang, Adam Q and Moulton, D and Lore, JD and Canik, J and Umansky, M and Wigram, Mike and Ballinger, S and Lipschultz, B and Bonnin, X},
  journal={Plasma Physics and Controlled Fusion},
  volume={65},
  number={3},
  pages={035011},
  year={2023},
  publisher={IOP Publishing}
}

@article{kotschenreuther2013magnetic,
  title={Magnetic geometry and physics of advanced divertors: The X-divertor and the snowflake},
  author={Kotschenreuther, Mike and Valanju, Prashant and Covele, Brent and Mahajan, Swadesh},
  journal={Physics of Plasmas},
  volume={20},
  number={10},
  year={2013},
  publisher={AIP Publishing}
}

@article{snyder2011first,
  title={A first-principles predictive model of the pedestal height and width: development, testing and ITER optimization with the EPED model},
  author={Snyder, PB and Groebner, RJ and Hughes, JW and Osborne, TH and Beurskens, M and Leonard, AW and Wilson, HR and Xu, XQ},
  journal={Nuclear Fusion},
  volume={51},
  number={10},
  pages={103016},
  year={2011}
}

@article{eich2017elm,
  title={ELM divertor peak energy fluence scaling to ITER with data from JET, MAST and ASDEX upgrade},
  author={Eich, Thomas and Sieglin, B and Thornton, AJ and Faitsch, M and Kirk, A and Herrmann, A and Suttrop, W and others},
  journal={Nuclear Materials and Energy},
  volume={12},
  pages={84--90},
  year={2017},
  publisher={Elsevier}
}

@inproceedings{dunne2025physics,
  title={The physics of ELM-free regimes in EUROfusion tokamaks; Pedestal tailoring via ballooning modes},
  author={Dunne, MG and Faitsch, M and Sauter, O and Viezzer, E and Labit, B and Kappatou, A and Keeling, D and JET, Contributors and ASDEX Upgrade Team and TCV Team and others},
  booktitle={30th IAEA Fusion Energy Conference (FEC 2025)},
  year={2025}
}

@article{greenwald1999characterization,
  title={Characterization of enhanced D$\alpha$ high-confinement modes in Alcator C-Mod},
  author={Greenwald, M and Boivin, R and Bonoli, P and Budny, R and Fiore, C and Goetz, J and Granetz, R and Hubbard, A and Hutchinson, I and Irby, J and others},
  journal={Physics of Plasmas},
  volume={6},
  number={5},
  pages={1943--1949},
  year={1999},
  publisher={American Institute of Physics}
}

@article{faitsch2021broadening,
  title={Broadening of the power fall-off length in a high density, high confinement H-mode regime in ASDEX Upgrade},
  author={Faitsch, Michael and Eich, T and Harrer, GF and Wolfrum, Elisabeth and Brida, D and David, P and Griener, M and Stroth, U and ASDEX Upgrade Team and EUROfusion MST1 Team and others},
  journal={Nuclear Materials and Energy},
  volume={26},
  pages={100890},
  year={2021},
  publisher={Elsevier}
}

@article{miller2025determination,
  title={Determination of confinement regime boundaries via separatrix parameters on Alcator C-Mod based on a model for interchange-drift-Alfv{\'e}n turbulence},
  author={Miller, Marco Andr{\'e}s and Hughes, Jerry W and Eich, Thomas and Tynan, George R and Manz, Peter and Body, T and Silvagni, Davide and Grover, Ond{\v{r}}ej and Hubbard, Amanda E and Cavallaro, A and others},
  journal={Nuclear Fusion},
  volume={65},
  number={5},
  pages={052002},
  year={2025},
  publisher={IOP Publishing}
}

@article{rogers1998phase,
  title={Phase space of tokamak edge turbulence, the L- H transition, and the formation of the edge pedestal},
  author={Rogers, BN and Drake, JF and Zeiler, A},
  journal={Physical Review Letters},
  volume={81},
  number={20},
  pages={4396},
  year={1998},
  publisher={APS}
}

@article{scott2005drift,
  title={Drift wave versus interchange turbulence in tokamak geometry: Linear versus nonlinear mode structure},
  author={Scott, Bruce D},
  journal={Physics of Plasmas},
  volume={12},
  number={6},
  year={2005},
  publisher={AIP Publishing}
}

@article{wolfrum2011characterization,
  title={Characterization of edge profiles and fluctuations in discharges with type-II and nitrogen-mitigated edge localized modes in ASDEX Upgrade},
  author={Wolfrum, Elisabeth and Bernert, M and Boom, JE and Burckhart, A and Classen, IGJ and Conway, GD and Eich, T and Fischer, R and Gude, A and Herrmann, A and others},
  journal={Plasma Physics and Controlled Fusion},
  volume={53},
  number={8},
  pages={085026},
  year={2011}
}

@article{harrer1292022,
  title={Quasicontinuous exhaust scenario for a fusion reactor: the renaissance of small edge localized modes},
  author={Harrer, GF and ASDEX Upgrade Team and EUROfusion MST1 Team and others},
  journal={Phys. Rev. Lett},
  volume={129},
  pages={165001},
  year={2022}
}

@article{creely2023sparc,
  title={SPARC as a platform to advance tokamak science},
  author={Creely, AJ and Brunner, D and Mumgaard, RT and Reinke, ML and Segal, M and Sorbom, BN and Greenwald, MJ},
  journal={Physics of Plasmas},
  volume={30},
  number={9},
  pages={090601},
  year={2023},
  publisher={AIP Publishing LLC}
}

@article{creely2020overview,
  title={Overview of the SPARC tokamak},
  author={Creely, AJ and Greenwald, Martin J and Ballinger, Sean B and Brunner, D and Canik, J and Doody, Jeffrey and F{\"u}l{\"o}p, T and Garnier, DT and Granetz, R and Gray, TK and others},
  journal={Journal of Plasma Physics},
  volume={86},
  number={5},
  pages={865860502},
  year={2020},
  publisher={Cambridge University Press}
}

@article{kuang2020divertor,
  title={Divertor heat flux challenge and mitigation in SPARC},
  author={Kuang, AQ and Ballinger, S and Brunner, D and Canik, John and Creely, AJ and Gray, Travis and Greenwald, M and Hughes, JW and Irby, J and LaBombard, B and others},
  journal={Journal of Plasma Physics},
  volume={86},
  number={5},
  pages={865860505},
  year={2020},
  publisher={Cambridge University Press}
}

@article{rodriguez2024core,
  title={Core performance predictions in projected SPARC first-campaign plasmas with nonlinear CGYRO},
  author={Rodriguez-Fernandez, P and Howard, NT and Saltzman, A and Shoji, L and Body, T and Battaglia, DJ and Hughes, JW and Candy, J and Staebler, GM and Creely, AJ},
  journal={Physics of Plasmas},
  volume={31},
  number={6},
  year={2024},
  publisher={AIP Publishing}
}

@article{wiesen2015new,
  title={The new SOLPS-ITER code package},
  author={Wiesen, Sven and Reiter, Detlev and Kotov, Vladislav and Baelmans, Martine and Dekeyser, Wouter and Kukushkin, Alexander S and Lisgo, Steven W and Pitts, Richard A and Rozhansky, Vladimir and Saibene, Gabriella and others},
  journal={Journal of nuclear materials},
  volume={463},
  pages={480--484},
  year={2015},
  publisher={Elsevier}
}

@article{Stangeby_2020_Part1,
  author    = {P. C. Stangeby},
  title     = {The roles of power loss and momentum-pressure loss in causing particle-detachment in tokamak divertors: I. A heuristic model analysis},
  journal   = {Plasma Physics and Controlled Fusion},
  volume    = {62},
  pages     = {025012},
  year      = {2020},
  month     = {dec},
  doi       = {10.1088/1361-6587/ab51a9},
  url       = {https://iopscience.iop.org/article/10.1088/1361-6587/ab51a9},
  publisher = {IOP Publishing}
}

@article{Stangeby_2020_Part2,
  author    = {P. C. Stangeby},
  title     = {The roles of power loss and momentum-pressure loss in causing particle-detachment in tokamak divertors: II. 2 Point Model analysis that includes recycle power-loss explicitly},
  journal   = {Plasma Physics and Controlled Fusion},
  volume    = {61},
  pages     = {025013},
  year      = {2020},
  month     = {dec},
  doi       = {10.1088/1361-6587/ab51d6},
  url       = {https://iopscience.iop.org/article/10.1088/1361-6587/ab51a9},
  publisher = {IOP Publishing}
}

@article{kotov2009two,
  title={Two-point analysis of the numerical modelling of detached divertor plasmas},
  author={Kotov, V and Reiter, D},
  journal={Plasma physics and controlled fusion},
  volume={51},
  number={11},
  pages={115002},
  year={2009}
}

@article{groth2015divertor,
  title={Divertor plasma conditions and neutral dynamics in horizontal and vertical divertor configurations in JET-ILW low confinement mode plasmas},
  author={Groth, M and Brezinsek, S and Belo, P and Brix, M and Calabro, G and Chankin, A and Clever, M and Coenen, JW and Corrigan, G and Drewelow, P and others},
  journal={Journal of nuclear materials},
  volume={463},
  pages={471--476},
  year={2015},
  publisher={Elsevier}
}

@article{moulton2018neutral,
  title={Neutral pathways and heat flux widths in vertical-and horizontal-target EDGE2D-EIRENE simulations of JET},
  author={Moulton, D and Corrigan, G and Harrison, JR and Lipschultz, B and JET Contributors},
  journal={Nuclear Fusion},
  volume={58},
  number={9},
  pages={096029},
  year={2018},
  publisher={IOP Publishing}
}

@article{eich2026ARC,
  title={Power and particle exhaust for the ARC fusion power plant},
  author={Eich, Thomas H and Body, Thomas AJ and Looby, Tom P and Ballinger, Sean B and Creely, Alexander J and Hillesheim, Jon C and Snyder, Philip B and Howard, Nathan T and Masline, Rebecca and Wigram, Michael RK and others},
  journal={Journal of Plasma Physics},
  volume={92},
  number={3},
  pages={E66},
  year={2026},
  publisher={Cambridge University Press}
}

\end{document}